\documentclass[12pt,journal,draftclsnofoot, onecolumn]{IEEEtran}

\usepackage{latexsym}
\usepackage[hidelinks]{hyperref}
\usepackage{tabularx}
\usepackage[T1]{fontenc}
\usepackage{algorithm}
\usepackage[noend]{algpseudocode}
\usepackage{graphicx}
\usepackage{amsmath,amssymb,amsfonts,stfloats}
\usepackage{array}
\usepackage{textcomp}
\usepackage[font=small]{caption}
\usepackage{paralist}
\usepackage{cite}
\usepackage{xcolor}
\usepackage{booktabs}
\usepackage[caption=false,font=footnotesize]{subfig}
\usepackage{url}
\usepackage{acronym}
\usepackage{latexsym}
\usepackage{hyperref}
\usepackage{soul}
\usepackage{multicol}
\usepackage{multirow}

\newcommand{\norm}[1]{\left\lVert#1\right\rVert}

\def\BibTeX{{\rm B\kern-.05em{\sc i\kern-.025em b}\kern-.08em T\kern-.1667em\lower.7ex\hbox{E}\kern-.125emX}}

\begin{document}

\title{Channel Estimation for 6G  V2X Hybrid Systems using Multi-Vehicular Learning}
\author{Marouan Mizmizi, Dario Tagliaferri , Damiano Badini, Christian Mazzucco, and Umberto Spagnolini
\thanks{Marouan Mizmizi, Dario Tagliaferri, and Umberto Spagnolini are with Dipartimento di Elettronica, Informazione e Bioingegneria, Politecnico di Milano.}
\thanks{Damiano Badini and Christian Mazzucco are with Huawei Technologies Italia}
\thanks{Umberto Spagnolini is also Huawei Industry Chair.}}

\maketitle

\begin{abstract}
\textbf{Channel estimation for hybrid Multiple Input Multiple Output (MIMO) systems at Millimeter-Waves (mmW)/sub-THz is a fundamental, despite challenging, prerequisite for an efficient design of hybrid MIMO precoding/combining. Most works propose sequential search algorithms, e.g., Compressive Sensing (CS), that are most suited to static channels and consequently cannot apply to highly dynamic scenarios such as Vehicle-to-Everything (V2X). To address the latter ones, we leverage \textit{recurrent vehicle passages} to design a novel Multi Vehicular (MV) hybrid MIMO channel estimation suited for Vehicle-to-Infrastructure (V2I) and Vehicle-to-Network (V2N) systems. Our approach derives the analog precoder/combiner through a MV beam alignment procedure. For the digital precoder/combiner, we adapt the Low-Rank (LR) channel estimation method to learn the position-dependent eigenmodes of the received digital signal (after beamforming), which is used to estimate the compressed channel in the communication phase.
Extensive numerical simulations, obtained with ray-tracing channel data and realistic vehicle trajectories, demonstrate the benefits of our solution in terms of both achievable Spectral Efficiency (SE) and Mean Square Error (MSE) compared to the Unconstrained Maximum Likelihood (U-ML) estimate of the compressed digital channel, making it suitable for both 5G and future 6G systems. Most notably, in some scenarios, we obtain the performance of the optimal Fully Digital (FD) systems.} 
\end{abstract}


\section{Introduction}
Recent advances in millimeter-wave (mmW) hardware \cite{8744505} and the potential availability of spectrum has encouraged the wireless industry to consider mmW, for the Fifth Generation of cellular systems (5G) \cite{5783993} and, in particular, for Vehicle-to-Everything (V2X) applications \cite{Garcia5GV2Xtutorial,8187182}. Following the same trend, sub-THz are envisioned for 6G systems \cite{Akyildiz2020_6G,Mogensen2020_6G,Wymeersch2021_6G}. Due to the increased carrier frequency, e.g., $24.25-52.6$ GHz for 5G New Radio (NR) Frequency Range 2 (FR2) and $>100$ GHz for sub-THz, mmW/sub-THz signals experience an orders-of-magnitude increase in free-space path loss compared to the current majority of wireless systems, resulting in highly sparse channels \cite{6834753,Akyildiz2015_subTHz}. Multiple Input Multiple Output (MIMO) systems are a redeeming solution that can provide a beamforming gain to overcome the path loss and establish links with a reasonable Signal-to-Noise Ratio (SNR). Additionally, MIMO systems enable precoding and combining of multiple data streams which could significantly improve the achievable data rate \cite{8169014,7244171}.

While the fundamental theory of MIMO precoding/combining is the same regardless of the carrier frequency, the hardware in the mmW/sub-THz band is subject to a set of non-trivial practical limitations. The processing in traditional MIMO systems is performed digitally at baseband, which requires a dedicated Radio Frequency (RF) chain for each antenna element. Unfortunately, due to the high number of elements required in mmW (even more at sub-THz), this implies a high cost and power consumption, which makes it unpractical \cite{8030501}.

A promising solution to these problems lies in the concept of hybrid arrays, which use a combination of analog beamforming in the RF domain and digital beamforming in the baseband, with a reduced number of RF chains. Hybrid Beamforming (HBF) was first introduced and analyzed in \cite{1519678}. It is driven by the fact that the number of RF chains is only lower-limited by the number of transmitted data streams, while the beamforming gain and diversity order is given by the number of antenna elements if proper precoding/combining is applied.
Analog precoding/combining is often implemented using phase shifters \cite{5754329}, \cite{7370753}, switches \cite{7394147}, or lenses \cite{7946172}. A HBF based on phase shifting network imposes the constraint of constant amplitude on the elements of the RF precoder. Moreover, there are two main HBF architectures, as shown in Fig. \ref{fig:BlockScheme}. On one hand, a Fully-Connected (FC-HBF) architecture, where each RF chain connects to all antenna elements of the array, while on the other hand, a Sub-Connected (SC-HBF) architecture, where the RF chains connect to disjoint subarrays, offering a cost-effective solution to HBF. Consequently, deriving the hybrid precoder/combiner is a complex, non-convex problem and therefore it is mathematically intractable \cite{6717211}.

\subsection*{Related Works}

Most  works  on  hybrid  precoding/combining  design \cite{6717211,8023810,7433949} require the knowledge of the full MIMO channel at both Transmitter (Tx) and Receiver (Rx). However, estimating the MIMO channel in mmW/sub-THz systems is a hard task due to the low Signal-to-Noise Ratio (SNR) before any beamforming. The presence of analog precoders/combiners implies that the digitally-observed channel is limited to a portion of  the full MIMO one, introducing an equivalent analog compression which cannot be handled with  conventional  channel  estimation  approaches \cite{7458188}. From the mathematical point of view, the channel decompression can be achieved by applying the hybrid echoing method proposed in \cite{7439748}, which consists of consecutively transmitting and receiving training sequences, while using all possible analog precoders/combiners (obtained, for example, as subset of a Fourier basis) and decompressing the channel after the concatenation of the received signals for each subset. However, this approach turns out to be infeasible for practical systems due to \textit{(i)} mobility of the terminals and \textit{(ii)} the low SNR resulting from mismatched Tx-Rx beams.

The authors in \cite{5262295, 6847111} propose a grid-based method for FC-HBF architecture, first estimating the Angles of Arrival/Departure (AoAs/AoDs) of the channel through a closed-loop beam training, after which the path gain of each pair AoA/AoD is derived. In \cite{8971952,8761521}, a similar approach is proposed for SC-HBF architecture under practical hardware impairments. In both architectures, the performance tends to be limited by the codebook resolution, while the complexity increases with the number of users. A different approach is based on Compressed Sensing (CS) techniques in \cite{5454399}, imposing a structured sparsity in the channel estimation problem. In \cite{7458188}, the CS-based open-loop approach is used to explicitly estimate the full channel, with a dictionary of quantized AoAs/AoDs. The results show the capability of CS to capture the full MIMO channel features allowing for the joint optimization of both analog and digital precoders/combiners. However, the algorithm requires an a-priori knowledge of the number of channel paths, and its performance is affected by the true sparsity level of the channel. Moreover, the joint optimization of both analog \textit{and} digital precoders/combiners increases the complexity and the cost of the implementation in practical high-mobility systems, as the channel is rapidly time-varying. Finally, as any grid-based technique, CS has a significant drawback in the high sensitivity to array calibrations \cite{Cerutti2020}, which is critical in hybrid systems \cite{8613274}.

Conversely, Low-Rank (LR) methods approach the MIMO channel estimation by exploiting the invariance of Spatial-Temporal (ST) channel features (i.e, AoA/AoD and delays) across different MIMO channel realizations, extracting a modal filtering on the received signal. LR are algebraic-based methods that leverage on the sparsity of the MIMO channel, as opposite to CS. Originally proposed in \cite{9149353,Nicoli2003,10.1109/TSP.2004.842191, 8553218} for low-frequency systems, where the channel is not sufficiently sparse to boost the LR application to practical systems, the LR has recently been studied for mmW/sub-THz systems, for Fully Digital (FD) systems only \cite{Cerutti2020,Nicoli2020}. In particular, the work in \cite{Cerutti2020} demonstrates that LR methods attain similar performances to CS with lower sensitivity to hardware impairments. In \cite{Cerutti2020}, the LR channel estimation is enabled by consecutive transmissions of training blocks, that limits the application to static or low-mobility scenarios.

\subsection*{Contribution}

In mobility, AoAs and AoDs describe an algebraic span of MIMO channel that has a LR, with a set of subspaces, and for mobile-to-fixed links both approaches are location dependent. Differently from the position-dependency of the of AoA/AoD in MIMO channel \cite{Nicoli2020}, here we first adapt this concept to Multi-Vehicular (MV) LR, and then we specialize the estimate to hybrid massive MIMO systems in mmW/sub-THz bands, considering both FC-HBF and SC-HBF architectures. Leveraging the algebraic properties of the MIMO channel and the constraints of hybrid hardware, we propose a two-stage training process for the algebraic estimation of single-user (e.g., or assuming that multi-users are allocated on minimally interfering angular, or frequency, or time radio resources), spatial hybrid MIMO channel in a mobile scenario. The proposed two-stage method applies to low mobility and high mobility scenarios, e.g., Vehicle-to-Infrastructure (V2I) or Vehicle-to-Network (V2N), which is the focus of this paper since it shows more challenging and interesting characteristics. In the first stage of pre-training (Section \ref{subsect:beamalignment}), we determine the optimal analog precoder/combiner at the Mobile Station (MS) and the Base Station (BS) (or Road Side Unit) through a MV codebook-based beam alignment procedure. In the second stage (Section \ref{subsect:LR}), the LR training system learns the algebraic channel subspace structure (eigenmodes) from the received training signal, observed at the digital side, used to obtain the LR-estimated channel. In particular, we propose two methods: an optimal approach, exploiting the \textit{joint} MS and BS spatial subspace, and another sub-optimal approach, considering the \textit{separate} MS and BS subspaces. Finally, during the communication phase (online phase), the MS and the BS use the pre-computed position-dependent analog precoder/combiner derived in the pre-training to transmit training sequences for digital LR estimation. Based on the so-called compressed channel (i.e., after analog BF \cite{7439748}), the BS derives the digital precoder and combiner. The novel aspect here is based on the position-dependency of either the analog precoders/combiners and the LR of the compressed MIMO channel obtained from \textit{multiple repeated vehicle passages} on the same geographical area, where the spatial features of the channel, i.e., AoAs/AoDs, are invariant. More specifically, the analog precoder/combiner and the LR eigenmodes are associated to the specific location in space of the MS. A notable advantage of the proposed approach is that, at the end of the training procedure, the BS stores a dataset of optimal analog precoders/combiners and digital channel eigenmodes, which does not require to be updated unless macroscopic changes in the environment occur. In this setting, during the communication phase the beam alignment can be avoided. This allows to reduce meaningfully the overall training overhead favoring the applications of the proposed channel estimation method to practical V2I/V2N systems. 

The proposed method is validated numerically, considering a realistic urban scenario and repeating-passage vehicle trajectories. The information on the building's geometry and road network topology is extracted from OpenStreetMap \cite{OpenStreetMap}, while, for vehicular mobility, we employ Simulator of Urban MObility (SUMO) software \cite{SUMO2018}. Finally, we use the ray-tracing software in \cite{WinProp} to generate the channel coefficients. Extensive simulations to assess the behavior of the proposed solution varying the channel (multipath vs. single-path), the SNR per antenna, the number of vehicles used during the training steps, and the number of RF chains (HBF setting). We observe that the proposed LR channel estimation method allows outperforming the Unconstrained Maximum Likelihood (U-ML) in terms of Mean Square Error (MSE) on channel estimation and Spectral Efficiency (SE) for both FC-HBF and SC-HBF architectures. In particular, for a target SE, both architectures and LR methods achieve an SNR gain up to 15 dB in single-path scenarios, and up to 10 dB in multipath scenarios. In general, the performance of the proposed channel estimation are proportional to the sparsity degree of the MIMO digital channel (after analog beamforming), which is high at mmW and it is even more prominent at sub-THz \cite{Akyildiz2015_subTHz}, making it suitable for 6G systems.

\subsection*{Organization}
The paper is organized as follows: Section \ref{sect:system_model} introduces the system and the channel model that are used throughout the paper. Section \ref{subsect:beamalignment} describes the proposed MV analog beam alignment, while Section \ref{subsect:LR} details the LR approach for hybrid MIMO systems. Section \ref{sect:results} reports the numerical results validating our work. Finally, Section \ref{sect:conclusion} draws the conclusions.

\subsection*{Notation}
Bold upper- and lower-case letters describe matrices and column vectors. $\left[\mathbf{A}\right]_{i,j}$ denotes the $(i,j)$ entry of matrix $\mathbf{A}$, while $\mathbf{A}^{(i)}$ is the $i$-th column. Matrix transposition and conjugate transposition is indicated as $(\cdot)^{\mathrm{T}}$ and $(\cdot)^{\mathrm{H}}$, respectively. $\norm{\cdot}$ denotes the Frobenius norm. $\mathrm{tr}\left(\mathbf{A}\right)$ and $\mathrm{rank}\left(\mathbf{A}\right)$ extracts the trace and the rank of matrix $\mathbf{A}$, respectively, while $\mathrm{eig}_{r}(\mathbf{A})$ is the collection of $r$ eigenvectors of $\mathbf{A}$. $\otimes$, $\diamond$ and $\odot$ denote, respectively, the Kronecker, the Kathri-Rao and the element-wise product between two matrices. $\mathrm{vec}(\cdot)$ denotes the vectorization by columns and $\mathrm{vec}^{-1}(\cdot)$ its inverse operation. $\mathrm{span}(\mathbf{A})$ denotes the subspace spanned by the columns of $\mathbf{A}$. $\mathbf{A}^{\dagger}$ is the Moore-Penrose pseudo-inverse of $\mathbf{A}$. $\mathrm{diag}(\cdot)$ denotes either a diagonal matrix or the extraction of the diagonal of a matrix. The following properties of the vectorization are used in the text: $\mathrm{vec}(\mathbf{A}\mathbf{B}\mathbf{C}) = (\mathbf{C}^{\mathrm{T}} \otimes \mathbf{A})\mathrm{vec}(\mathbf{B})$, $\mathrm{vec}(\mathbf{A}\mathbf{B}) = (\mathbf{B}^{\mathrm{T}} \otimes \mathbf{I})\mathrm{vec}(\mathbf{A})$. With  $\mathbf{a}\sim\mathcal{CN}(\boldsymbol{\mu},\mathbf{C})$ we denote a multi-variate complex Gaussian random variable $\mathbf{a}$ with mean $\boldsymbol{\mu}$ and covariance $\mathbf{C}$. $\mathbb{E}[\cdot]$ is the expectation operator, while $\mathbb{R}$ and $\mathbb{C}$ stand for the set of real and complex numbers, respectively. $\delta_{n}$ is the Kronecker delta.

\section{System and Channel Model}\label{sect:system_model}

\begin{figure}[t!]
\centering
    \subfloat[][FC-HBF Tx]{\includegraphics[width=.48\columnwidth]{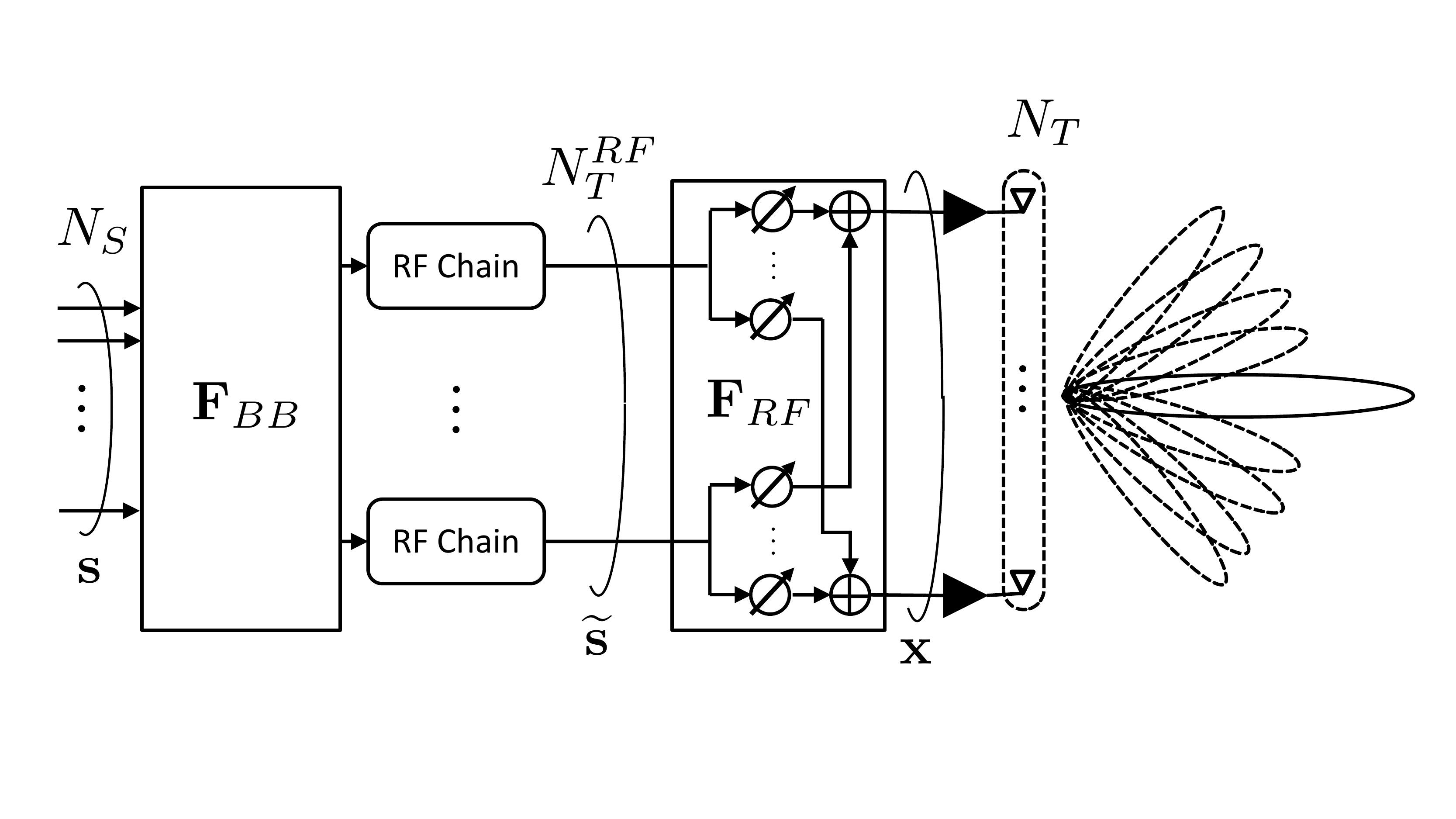}\label{subfig:Txfully}} \hspace{0.2mm}
    \subfloat[][FC-HBF Rx]{\includegraphics[width=.48\columnwidth]{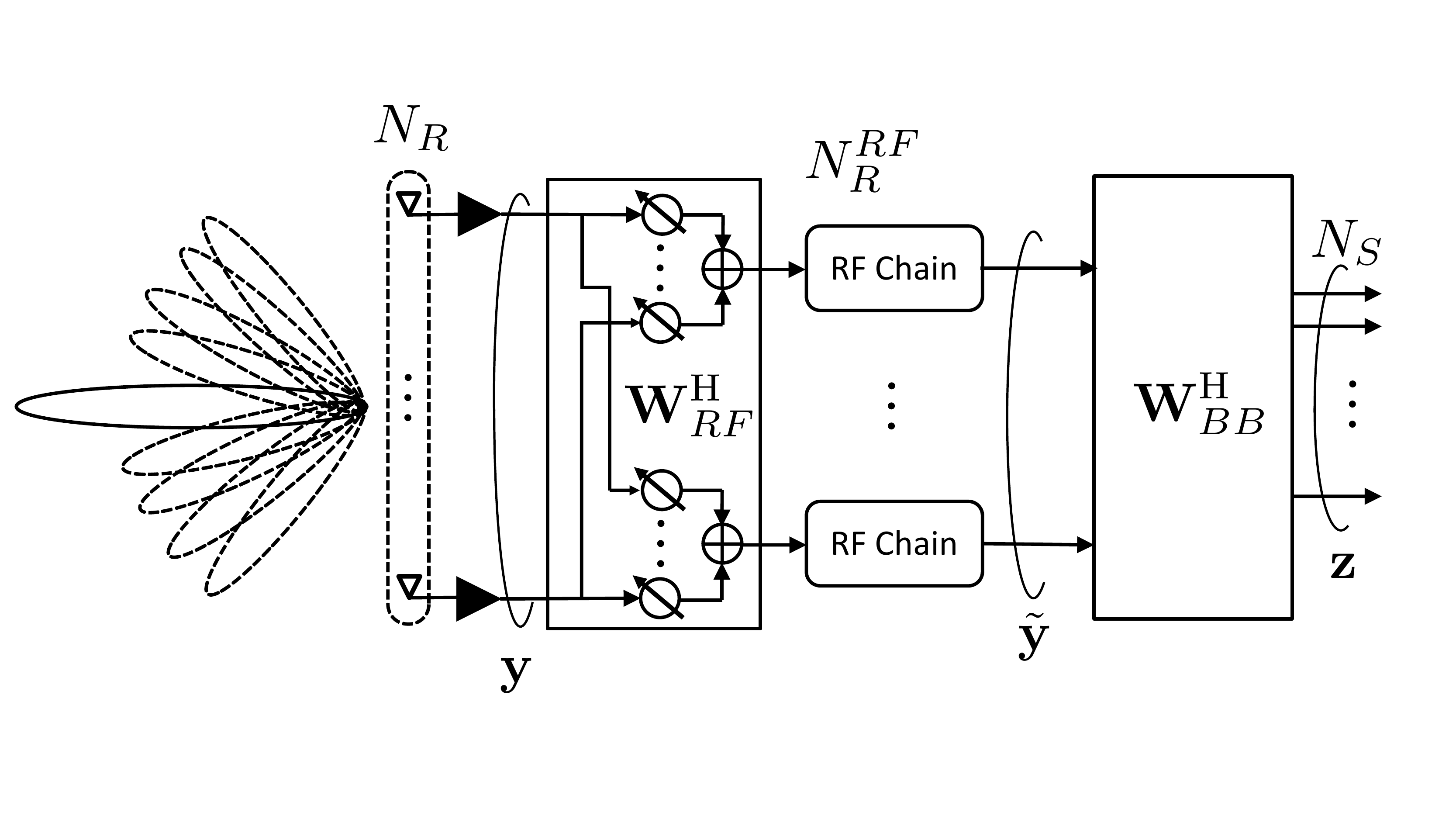}\label{subfig:Rxfully}}\\ 
    \subfloat[][SC-HBF Tx]{\includegraphics[width=.48\columnwidth]{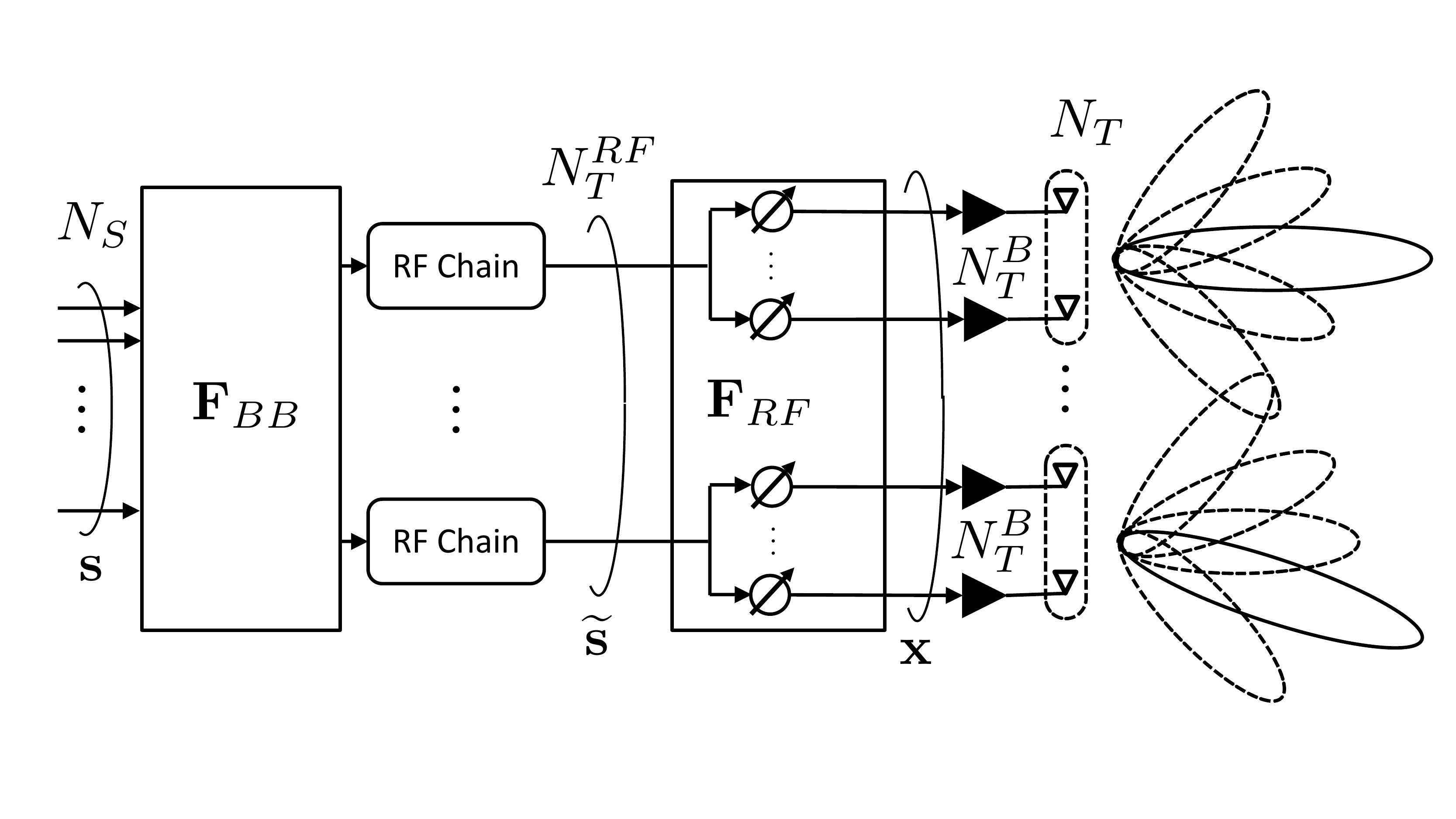}\label{subfig:Txsub}} \hspace{0.2mm}
    \subfloat[][SC-HBF Rx]{\includegraphics[width=.48\columnwidth]{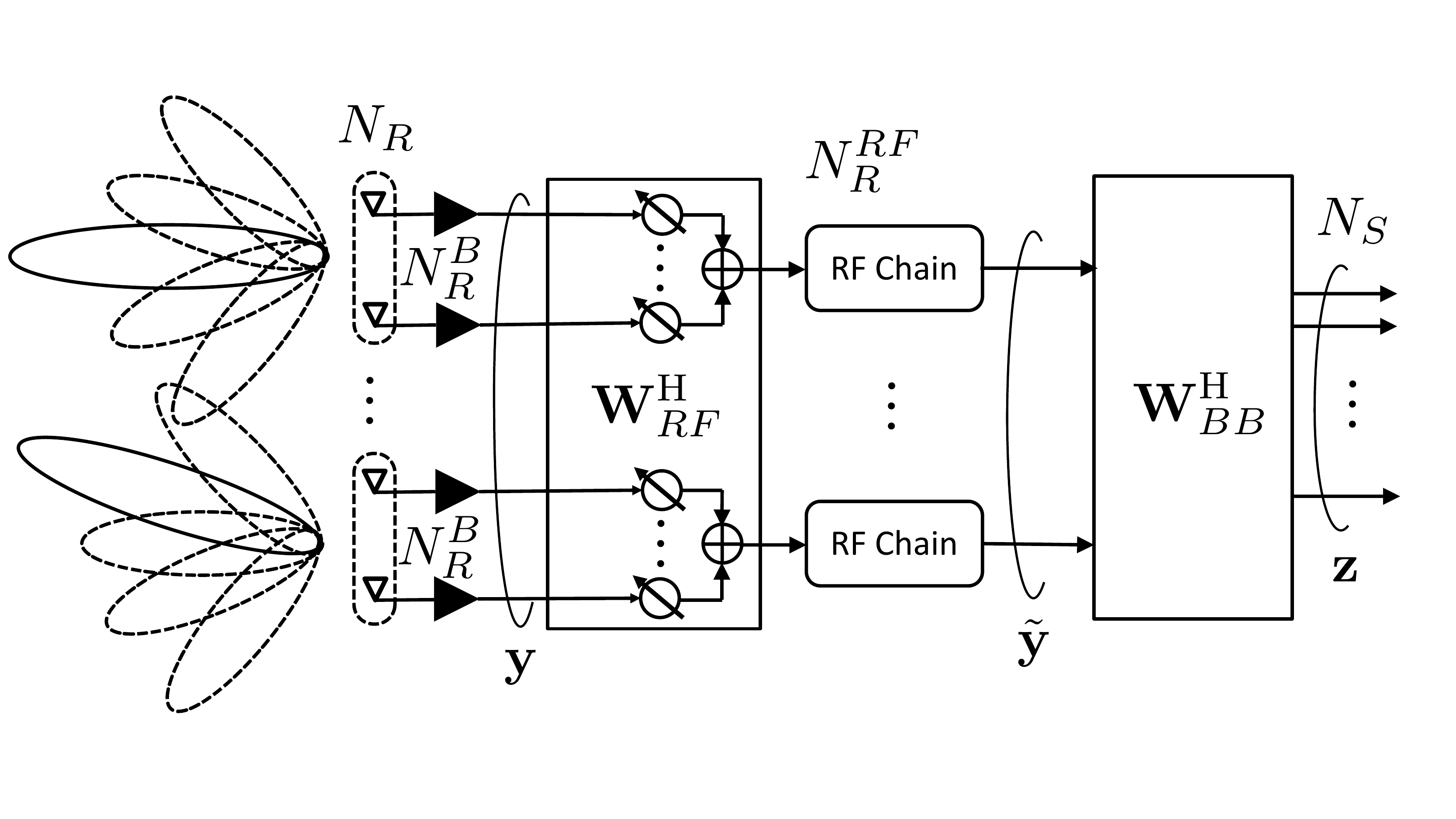}\label{subfig:Rxsub}}\\
    \caption{Block scheme of the FC-HBF (\ref{subfig:Txfully},\ref{subfig:Rxfully}) and SC-HBF (\ref{subfig:Txsub},\ref{subfig:Rxsub}) hybrid MIMO system}
    \label{fig:BlockScheme}
\end{figure}

We consider the single-user hybrid mmWave MIMO system depicted in Fig. \ref{fig:BlockScheme}. The Tx is equipped with $N_T$ antenna elements and $N_T^{RF}$ RF chains, that is communicating $N_S$ data streams. The Rx has $N_R$ antenna elements and $N^{RF}_R$ RF chains. The hybrid hardware configuration consists in $N_T^{RF} < N_T$ and $N_R^{RF} < N_R$, while the number of parallel data streams $N_S$ is upper-bounded as $N_S\leq\mathrm{min}(N_R^{RF}, N_T^{RF})$. For the sub-connected configuration, the Tx and Rx antennas are grouped into sub-arrays of $N_T^B$ and $N_R^B$ antennas, respectively, each one connected to a single RF chain, i.e., $N_T^B = N_T/N_T^{RF}$ and $N_R^B = N_R/N_R^{RF}$. 
The $N_S$ complex symbols to be transmitted are $\mathbf{s} \in \mathbb{C}^{N_S\times 1}\sim \mathcal{CN}\left(\mathbf{0}, \mathbf{I}_{N_S}/N_S\right)$, and are precoded using the cascade of $\mathbf{F}_{BB} \in \mathbb{C}^{N^{RF}_{T} \times N_S}$, obtaining the digital signal vector $\widetilde{\mathbf{s}} = \mathbf{F}_{BB}\, \mathbf{s} \in\mathbb{C}^{N_T^{RF}\times 1}$, and of $\mathbf{F}_{RF} \in \mathbb{C}^{N_{T}\times N^{RF}_{T}}$ in the analog domain. The discrete-time transmitted signal is therefore given by:
\begin{equation}\label{eq:Analog&DigitalPrec}
    \mathbf{x} = \mathbf{F}_{RF}\; \widetilde{\mathbf{s}},
\end{equation}
where $\mathbf{x}\in\mathbb{C}^{N_T \times 1}$. For channel estimation, an orthogonal training sequence $\mathbf{v}\in\mathbb{C}^{N_T^{RF}\times 1}$, detailed in Section \ref{subsect:LR}, is transmitted without the digital precoder $\mathbf{F}_{BB}$, i.e., $\widetilde{\mathbf{s}} = \mathbf{v}$.

Since $\mathbf{F}_{RF}$ is implemented using analog phased shifters, its elements are constrained to have the same norm, i.e., $[\mathbf{F}^{(i)}_{RF}\mathbf{F}^{(i),\mathrm{H}}_{RF}]_{k,k} = 1/N_T$,
while the Tx total power constraint is enforced by designing $\mathbf{F}_{BB}$ such that $\norm{\mathbf{F}_{RF}\mathbf{F}_{BB}}^2 = N_S$. In the SC-HBF configuration, the $N_T^B N_T^{RF}\times N_T^{RF}$ analog precoding matrix $\mathbf{F}_{RF}$ is block-diagonal:
\begin{equation}\label{eq:Frf_subconnected}
    \mathbf{F}_{RF} = \begin{bmatrix}\mathbf{f}^{(1)}_{RF} & \mathbf{0} &\cdots & \mathbf{0}\\
    \mathbf{0} & \mathbf{f}^{(2)}_{RF} & \cdots & \mathbf{0}\\
    \vdots & \vdots & \vdots & \vdots\\
    \mathbf{0} &  \cdots & \mathbf{0} & \mathbf{f}^{(N_T^{RF})}_{RF}\end{bmatrix},
\end{equation}
where $\mathbf{f}^{(n)}_{RF}\in\mathbb{C}^{N_T^B \times 1}$, $n=1,\dots,N_T^{RF}$ is the beamforming vector for the $n$-th Tx sub-array.

The transmitted signal is assumed to propagate in a spatially-sparse channel $\mathbf{H} \in \mathbb{C}^{N_R \times N_T}$ affected, for simplicity, by block-fading \cite{6717211}. After the time-frequency synchronization, the received signal is:
\begin{equation}\label{eq:propagation}
    \mathbf{y} = \mathbf{H} \mathbf{x} + \mathbf{n}
\end{equation}
where noise $\mathbf{n} \sim \mathcal{CN}\left(\mathbf{0}, \mathbf{Q}_n\right)$ is generally not white due to the presence of directional interference. Similarly to the Tx, the Rx applies the cascade of analog and digital combiners, here indicated with $\mathbf{W}_{RF} \in \mathbb{C}^{N_{R}\times N^{RF}_{R}}$ and $\mathbf{W}_{BB} \in \mathbb{C}^{N_{R}^{RF}\times N_S}$, respectively. The (compressed) digital signal $\widetilde{\mathbf{y}}\in\mathbb{C}^{N_R^{RF} \times 1}$ after the analog combiner $\mathbf{W}_{RF}$ is:
\begin{equation}\label{eq:AnalogComb}
\begin{split}
    \widetilde{\mathbf{y}} = \underbrace{\mathbf{W}^\mathrm{H}_{RF} \mathbf{H} \mathbf{F}_{RF}}_{\widetilde{\mathbf{H}}}\, \widetilde{\mathbf{s}} +  \widetilde{\mathbf{n}}
\end{split}
\end{equation}
where: 
\begin{itemize}
    \item $\mathbf{W}_{RF}$ compressing the analog signal is subject to the same constraint of $\mathbf{F}_{RF}$, i.e., $[\mathbf{W}^{(j)}_{RF}\mathbf{W}^{(j),\mathrm{H}}_{RF}]_{l,l} = 1/N_R$;
    \item $\widetilde{\mathbf{H}}\in\mathbb{C}^{N_R^{RF}\times N_T^{RF}}$ is the equivalent and compressed MIMO channel observed at the digital side;
    \item the noise \textit{after} the analog beamforming is $\widetilde{\mathbf{n}} = \mathbf{W}^\mathrm{H}_{RF}\,\mathbf{n}\sim \mathcal{CN}(\mathbf{0}, \widetilde{\mathbf{Q}}_n)$, with $\widetilde{\mathbf{Q}}_n = \mathbf{W}^\mathrm{H}_{RF} \mathbf{Q}_n \mathbf{W}ì_{RF}$.
\end{itemize}
Similarly to $\mathbf{F}_{RF}$, the analog combiner $\mathbf{W}_{RF}$ for SC-HBF architectures is block-diagonal. 

Finally, the received data flows $\mathbf{z}\in\mathbb{C}^{N_S\times 1}$ after the digital combiner $\mathbf{W}_{BB}$ are:
\begin{equation}\label{eq:Analog&DigitalComb}
\begin{split}
    \mathbf{z} = \mathbf{W}^{\mathrm{H}}_{BB} \widetilde{\mathbf{y}} = \mathbf{W}^{\mathrm{H}}_{BB} \widetilde{\mathbf{H}}\, \widetilde{\mathbf{s}} + \mathbf{W}^{\mathrm{H}}_{BB} \widetilde{\mathbf{n}}.
\end{split}
\end{equation}
%
Derivation of $\mathbf{F}_{BB}, \mathbf{F}_{RF}, \mathbf{W}_{BB}, \mathbf{W}_{RF}$ has been investigated in depth in \cite{6717211}. Here, the analog precoders/combiners $\mathbf{F}_{RF}$ and $\mathbf{W}_{RF}$ are derived from a MV codebook-based beam alignment procedure. After, the digital precoders $\mathbf{F}_{BB}$ and combiners $\mathbf{W}_{BB}$ are computed employing the LR training in the second stage, as detailed in Section \ref{subsect:LR}. The aforementioned system model refers, for instance, to one sub-carrier of an OFDM system and temporal processing over the sub-carriers is not detailed herein.

\subsection{Channel Model}\label{subsect:channel_model}

As customary in mmW/sub-THz links, we consider the spatially-sparse clustered MIMO channel model \cite{995521, 1033686}. The channel matrix $\mathbf{H}$ can be written as the sum of $P$ paths as
\begin{equation}\label{eq:channel_matrix}
    \mathbf{H} = \sum_{p=1}^{P}\alpha_p\,\mathbf{a}_R(\boldsymbol{\vartheta}_p)\mathbf{a}^{\mathrm{T}}_T(\boldsymbol{\psi}_p),
\end{equation}
where: \textit{(i)} $\alpha_p$ is the complex gain of the $p$-th path; \textit{(ii)} $\mathbf{a}_T(\boldsymbol{\psi}_p) \in \mathbb{C}^{N_T\times 1}$ and $\mathbf{a}_R(\boldsymbol{\vartheta}_p) \in \mathbb{C}^{N_R\times 1}$  represent, respectively, the Tx and Rx and array response vectors to $p$-th path, function of the AoDs $\boldsymbol{\psi}_p = [\psi_{\mathrm{az},p}, \psi_{\mathrm{el},p}]^\mathrm{T}$ and the AoAs $\boldsymbol{\vartheta}_p = [\vartheta_{\mathrm{az},p}, \vartheta_{\mathrm{el},p}]^\mathrm{T}$.

Without loss of generality, we assume the faded channel to be normalized such that $\mathbb{E}[\norm{\mathbf{H}}^2] = N_T N_R$. The channel matrix \eqref{eq:channel_matrix} can be rewritten in compact form as:
\begin{equation}\label{eq:channel_matrix_compact}
    \mathbf{H} = \mathbf{A}_R\left(\boldsymbol{\vartheta}\right) \mathbf{D}\,\mathbf{A}_T^{\mathrm{T}}\left(\boldsymbol{\psi}\right),
\end{equation}
where $\mathbf{A}_T\left(\boldsymbol{\psi}\right) = \left[\mathbf{a}_T(\boldsymbol{\psi}_1),\dots,\mathbf{a}_T(\boldsymbol{\psi}_P)\right] \in \mathbb{C}^{N_T\times P}$ and
$\mathbf{A}_R\left(\boldsymbol{\vartheta}\right) = \left[\mathbf{a}_R(\boldsymbol{\vartheta}_1),\dots,\mathbf{a}_R(\boldsymbol{\vartheta}_P)\right] \in \mathbb{C}^{N_R\times P}$ are two matrices identifying the Tx and Rx \textit{beam spaces}, and diagonal matrix $\mathbf{D} \in \mathbb{C}^{P\times P}= \mathrm{diag}\left(\alpha_1, \dots, \alpha_P\right)$ collects all the channel amplitudes, obeying the Wide-Sense Stationary Uncorrelated Scattering (WSSUS) model \cite{661053}:
\begin{equation}\label{eq:scatteringamplitude_corr}
    \mathbb{E}\left[ \mathbf{D}_n \mathbf{D} ^{\mathrm{H}}_{n+m}\right] = \mathbf{P}\delta_{n-m},
\end{equation}
with $\mathbf{P} = \mathrm{diag}\left(P_1, \dots,P_P\right)$ containing the paths' powers, normalized such that $\sum_p P_p = 1$, and $n$, $m$ denoting two different channel realizations in either time (different fading blocks) or space (different locations).

Matrices $\mathbf{A}_T\left(\boldsymbol{\psi}\right)$ and $\mathbf{A}_R\left(\boldsymbol{\vartheta}\right)$ allow to define the diversity orders of channel $\mathbf{H}$ for Tx ($r_{T}$) and Rx ($r_{R}$) 
\begin{align}\label{eq:TX_RX_diversity_orders}
    r_{T} & = \mathrm{rank}(\mathbf{A}_T\left(\boldsymbol{\psi}\right)) \leq \mathrm{min}\left(N_T, P\right),\\
    r_{R} & = \mathrm{rank}(\mathbf{A}_R\left(\boldsymbol{\vartheta}\right))  \leq \mathrm{min}\left(N_R, P\right),
\end{align}
i.e., the number of resolvable spatial paths according to the number of Tx and Rx antennas. 

The analog precoder/combiner pair $\mathbf{F}_{RF}$ and $\mathbf{W}_{RF}$ modifies the beam spaces and the diversity orders of the digitally-equivalent channel $\widetilde{\mathbf{H}}$, whose structure is:
\begin{equation}\label{eq:compressed_channel_matrix}
    \widetilde{\mathbf{H}} = \mathbf{W}^{\mathrm{H}}_{RF} \mathbf{A}_R(\boldsymbol{\vartheta}) \, \mathbf{D}\, \mathbf{A}_T^{\mathrm{T}}(\boldsymbol{\psi})\mathbf{F}_{RF}.
\end{equation}
The diversity orders, namely number of resolvable paths given the Tx and RX HBF configurations, now become:
\begin{align}
    &\widetilde{r}_{T} =  \mathrm{rank}(\mathbf{A}_T^{\mathrm{T}}(\boldsymbol{\psi})\mathbf{F}_{RF}) \leq \mathrm{min}\left(\mathrm{rank}(\mathbf{F}_{RF}),r_T\right), \label{eq:Req_rank_DS_Tx}\\
    &\widetilde{r}_{R} =  \mathrm{rank}(\mathbf{W}^{\mathrm{H}}_{RF} \mathbf{A}_R(\boldsymbol{\vartheta})) \leq \mathrm{min}\left(\mathrm{rank}(\mathbf{W}_{RF}),r_R\right),  \label{eq:Req_rank_DS_Rx}
\end{align}
where, in general, 
\begin{align}\label{eq:analog_precodercombiner_rank}
    \mathrm{rank}(\mathbf{F}_{RF}) \leq N_T^{RF}, \,\,\,
    \mathrm{rank}(\mathbf{W}_{RF}) \leq N_R^{RF}.
\end{align}
As will be shown in the following, as the full MIMO channel cannot be directly estimated, \cite{6717211,5262295}, we exploit the algebraic structure of the digitally-observed channel $\widetilde{\mathbf{H}}$ to improve the channel estimation. 
As opposite to existing works \cite{5262295, 7439748, 6847111, 7458188}, we propose a learning-based approach, tailored for both static and dynamic scenarios (e.g., V2I/V2N scenarios). A MV codebook-based beam alignment procedure selects the analog precoder/combiner pair $\mathbf{F}_{RF}$, $\mathbf{W}_{RF}$ (Section \ref{subsect:beamalignment}); then, a second MV-LR method learns the algebraic spatial eigenmodes of the digital compressed channel $\widetilde{\mathbf{H}}$, which are used to derive digital precoders/combiners $\mathbf{F}_{BB}$ and $\mathbf{W}_{BB}$ from the LR-estimated equivalent compressed channel (Section \ref{subsect:LR}).

\section{Multi-vehicular Codebook-based Analog Beam Alignment} \label{subsect:beamalignment}

The hardware constraint and the low SNR in the mmW/sub-THz bands makes the derivation of analog precoder/combiner in hybrid systems is a complex non-convex problem \cite{7990158}. A conventional solution is to use a fixed codebook and a beam alignment strategy to appropriately scan the full channel (both AoA and AoD) and to select the best beam pairs that satisfy some criterion, such as to maximize the received power, the SNR, or the achievable rate. The trade-off between complexity and resolution must be taken into account when designing the codebook \cite{7947209}. We elaborate further below from this beam-alignment approach.

\begin{figure}
    \centering
    \subfloat[][]{\includegraphics[width=.48\columnwidth]{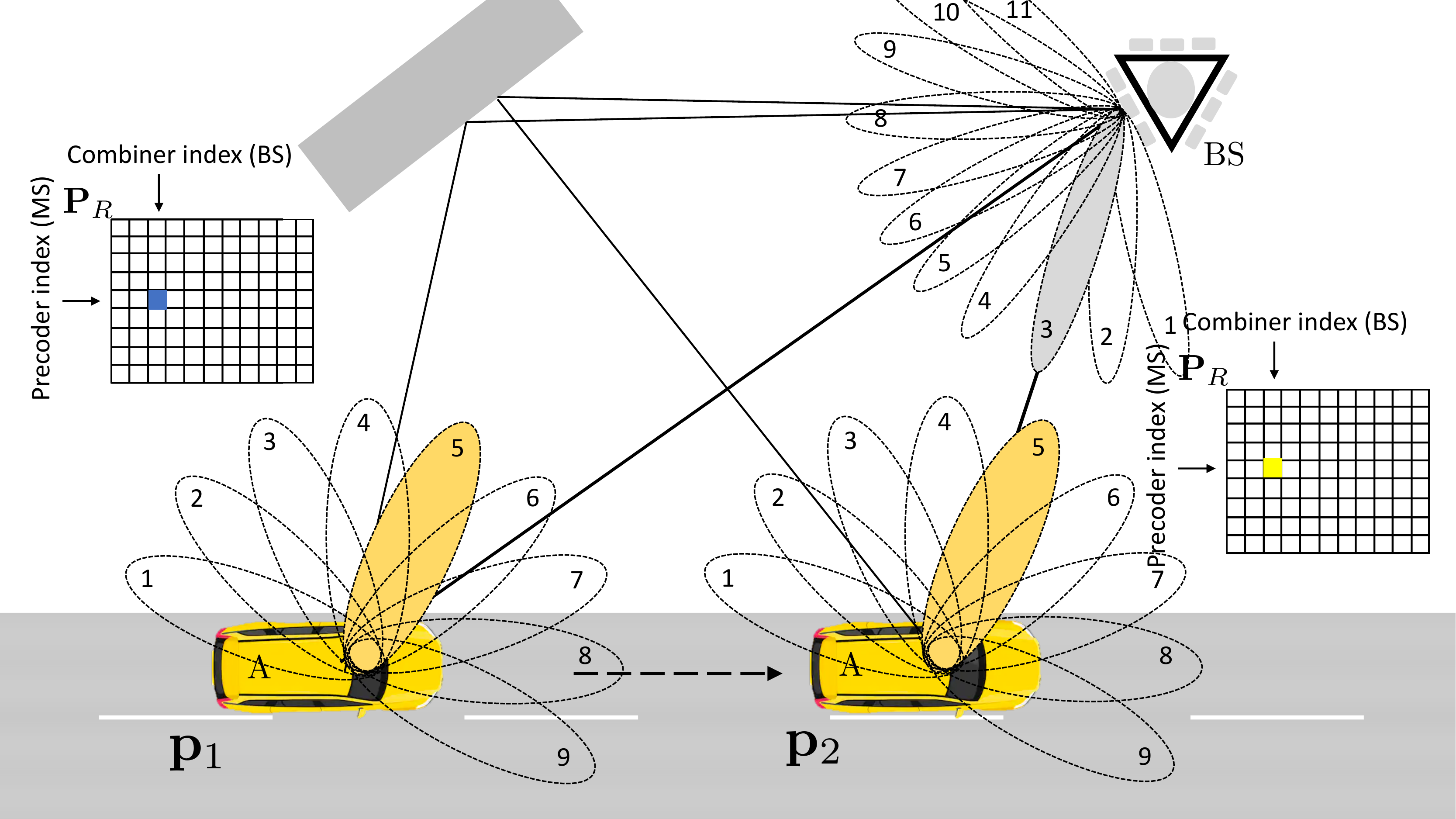}\label{subfig:MVclustering1}}\hspace{0.2mm}
    \subfloat[][]{\includegraphics[width=.48\columnwidth]{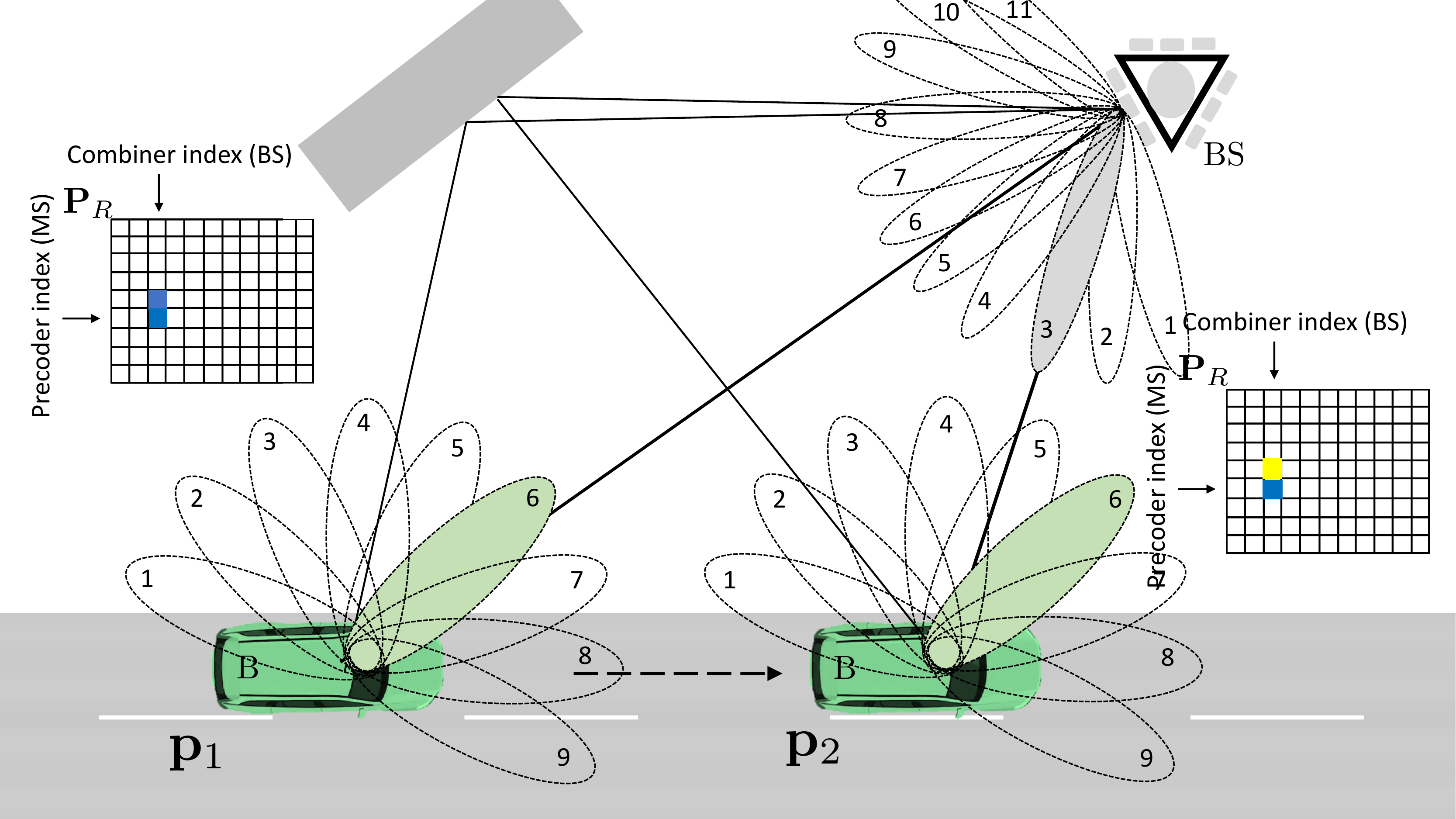}\label{subfig:MVclustering2}}\\
    \subfloat[][]{\includegraphics[width=.48\columnwidth]{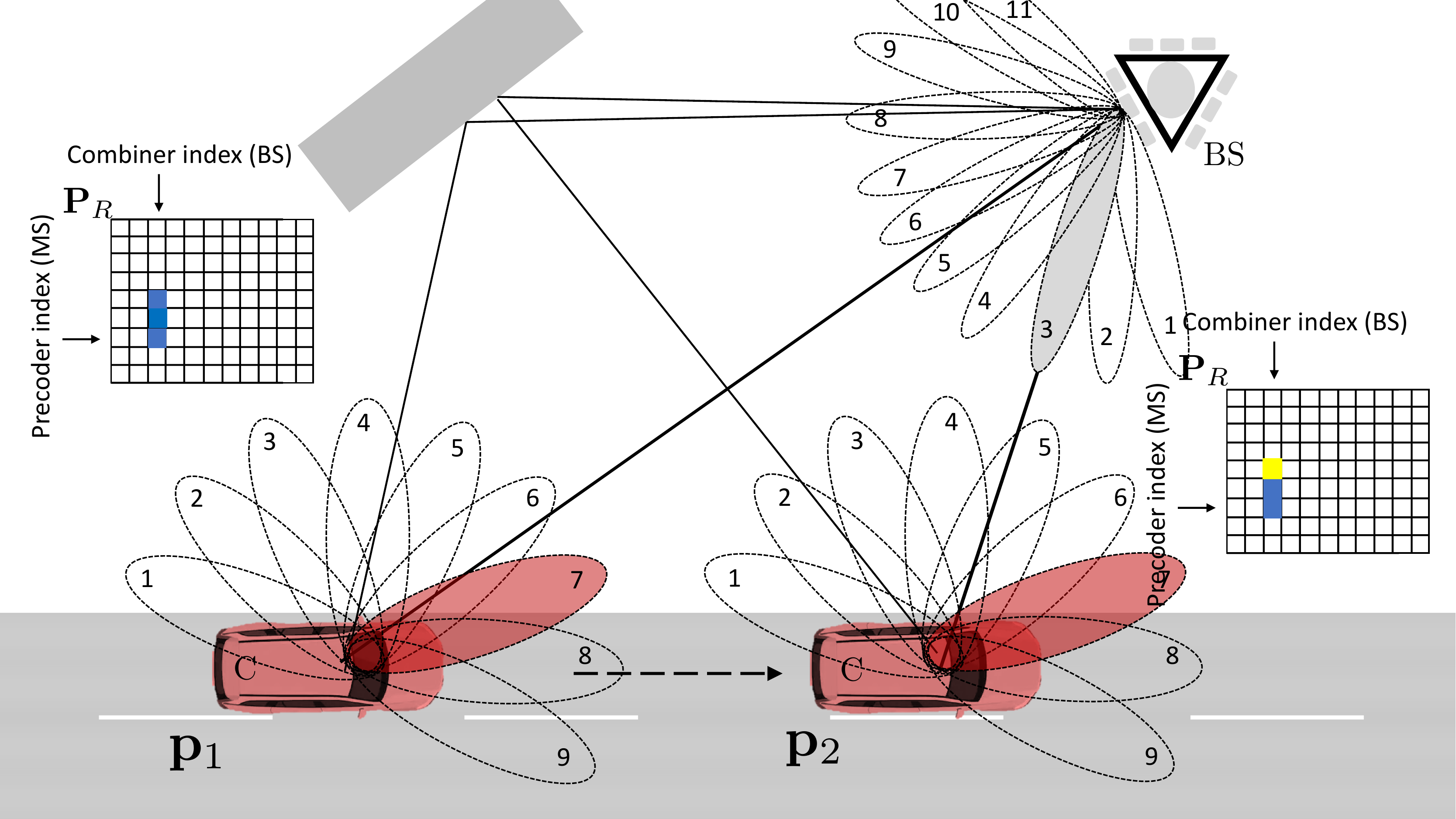}\label{subfig:MVclustering3}}\hspace{0.2mm}
    \subfloat[][]{\includegraphics[width=.48\columnwidth]{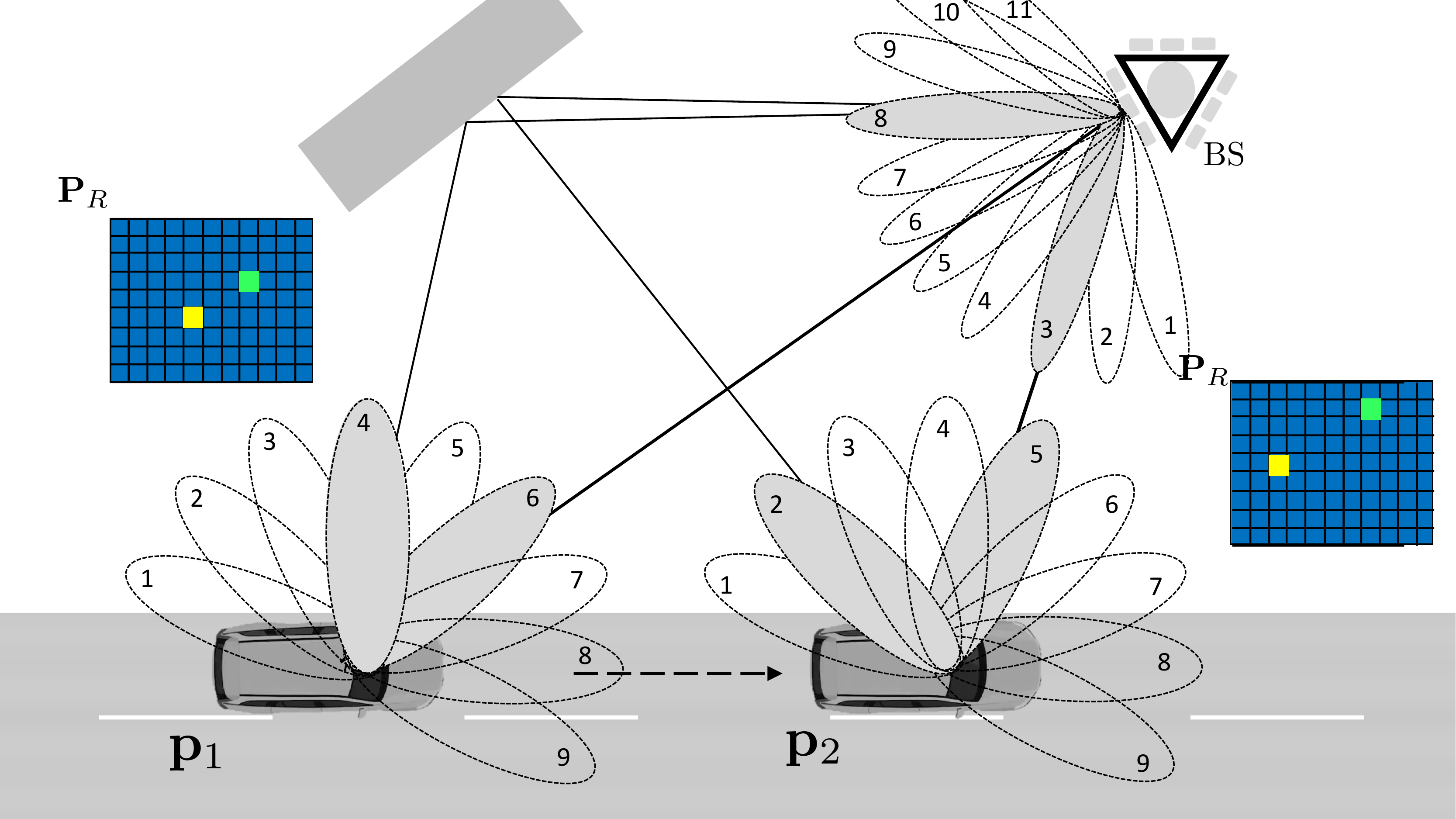}\label{subfig:MVclustering4}}\\
    \caption{Graphical representation of the MV codebook-based beam alignment: the BS assigns different analog beams to different MSs (\ref{subfig:MVclustering1},\ref{subfig:MVclustering2},\ref{subfig:MVclustering3}), collecting a set of position-related power measurements used to progressively filling matrix $\mathbf{P}_R$. At the end of the procedure, the MV beam alignment selects the best analog beam pairs (i.e., analog precoder/combiner pair) (\ref{subfig:MVclustering4})}
    \label{fig:MVbeamalignment}
\end{figure}

In a quasi-static propagation environment, different vehicles crossing the same location in space with slightly different co-directed trajectories (as commonly happens in urban scenarios) experience the same angles (AoD/AoA) in communicating with the BS and different fading amplitudes made varying by the Doppler \cite{Chang8594703}. Therefore, by leveraging this property, we explore a set of MIMO channel snapshots of \textit{recurrent vehicle passages} to design a novel \textit{multi-vehicular} codebook-based analog beam alignment procedure for dynamic scenarios, characterized by the mobility of one of the terminals, e.g., in V2I/V2N communications. We assume the Rx, e.g., a BS, with a fixed position and a set of \textit{collaborative} vehicles, both equipped with hybrid antenna arrays. In particular, all the MSs have the same array equipment, and their positions $\mathbf{p}_{\ell}$ and headings $\theta_{\ell}$ are known, with a reasonable accuracy, for each training block ($\ell$-th MS). The way the position and orientation are obtained is out of the scope but one can use \textit{(i)} a Radio Access Technology (RAT)-based localization and tracking algorithms \cite{8690640}, or \textit{(ii)} some signaling from the vehicle's onboard sensors (e.g., Global Navigation Satellite System (GNSS)). 

The Tx analog codebook is designed from a 2D Fourier basis that for a $N_1 \times N_2$ Uniform Rectangular Array (URA) with half-wavelength spaced antennas configuration becomes:
\begin{equation}\label{eq:2Dcodebook}
    \mathbf{B}_{\mathrm{2D}}\left(N_1,N_2\right) = \mathbf{B}\left(N_1\right) \otimes \mathbf{B}\left(N_2\right),
\end{equation}
where $\mathbf{B}\left(N\right)\in\mathbb{C}^{N \times N}$ is the DFT matrix with entries
\begin{equation}
    \left[\mathbf{B}\left(N\right)\right]_{m,n} = \frac{1}{\sqrt{N}} e^{-j \frac{2 \pi m n }{N}},
\end{equation}
and dimensions are
\begin{equation}
    \begin{split}
           & N_1 = N_T^{\mathrm{az}}, \,\,\,\, N_2 = N_T^{\mathrm{el}} \quad \text{for FC-HBF},\\
           & N_1 = N_T^{B,\mathrm{az}}, \,\, N_2 = N_T^{B,\mathrm{el}} \quad \text{for SC-HBF}
    \end{split}
\end{equation}
in which $N_T^{\mathrm{az}}$, $N_T^{\mathrm{el}}$ denote the number of Tx antennas along the azimuth (horizontal) and elevation (vertical) direction of the URA ($N_T = N_T^{\mathrm{az}} \times N_T^{\mathrm{el}}$), and $N_T^{B,\mathrm{az}}$, $N_T^{B,\mathrm{el}}$ the same for each sub-array ($N_T^B = N_T^{B,\mathrm{az}} \times N_T^{B,\mathrm{el}}$). The Rx codebook is analogously obtained.

\begin{figure}[!b]
    \centering
    \subfloat[FC-HBF]{\includegraphics[width=0.48\columnwidth]{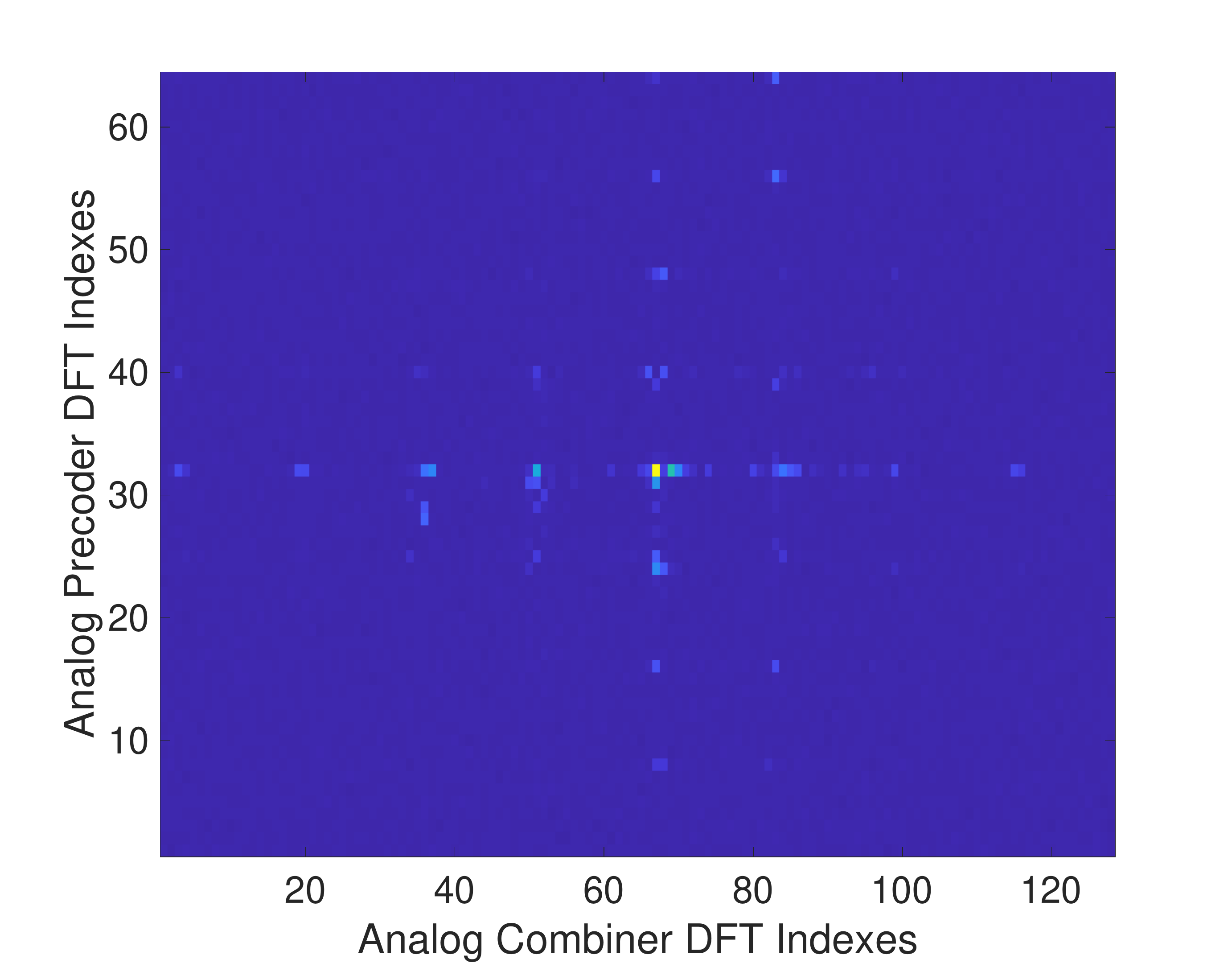}\label{subfig:PRfully}}\hspace{0.2mm}
    \subfloat[SC-HBF]{\includegraphics[width=0.48\columnwidth]{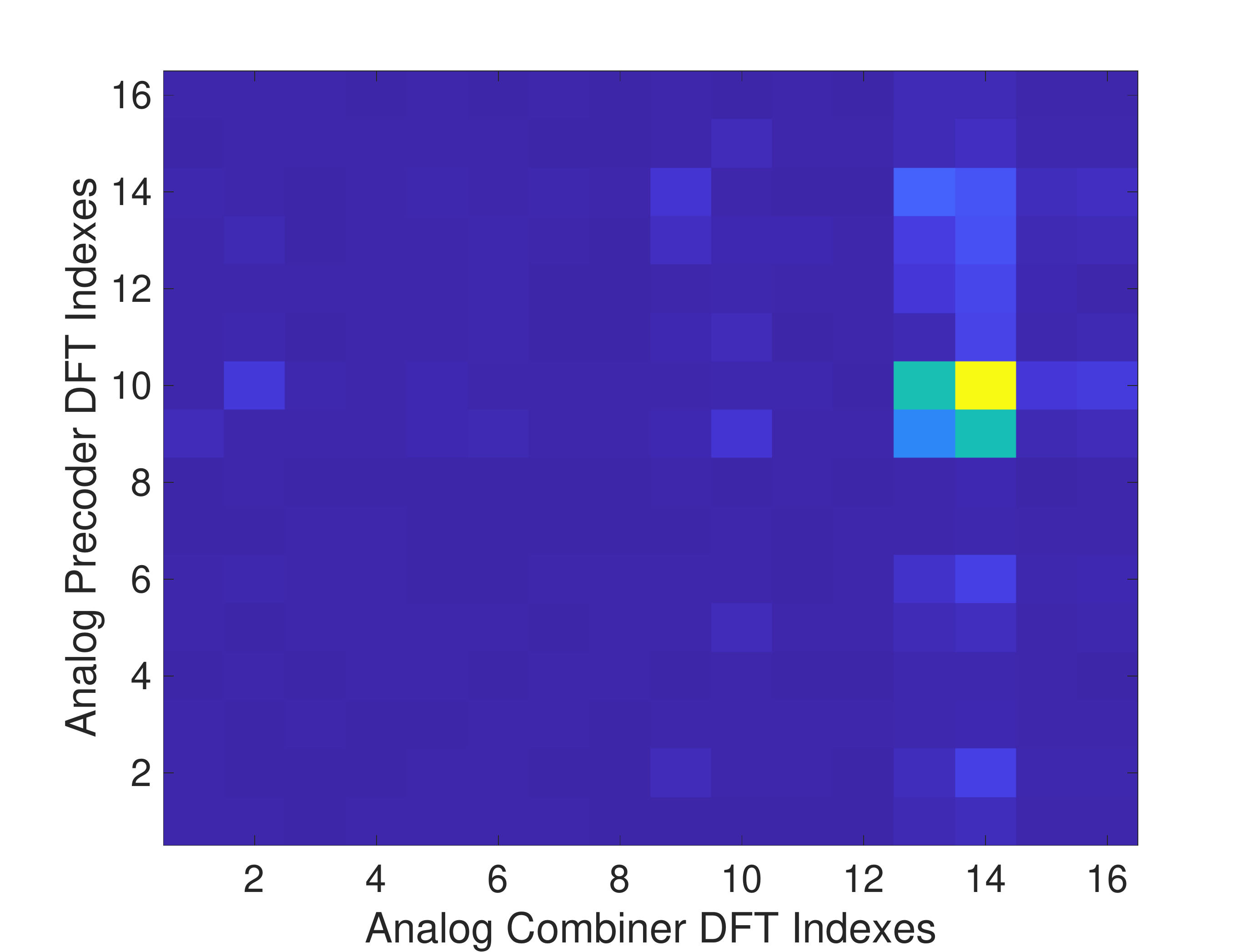}\label{subfig:PRsub}}\\
    \caption{Example of $\mathbf{P}_{R}$, with $N_T=64$, $N_R=128$ and $N^{RF}_T=4$, $N^{RF}_R=8$ hybrid array configuration, for FC-HBF (\ref{subfig:PRfully}) and SC-HBF (\ref{subfig:PRsub}) architecture}
    \label{fig:ExamplePR}
\end{figure}

The learning stage of the MV codebook-based beam alignment procedure is depicted in Fig. \ref{fig:MVbeamalignment} and it consists on the following steps, that involve the usage of a low-frequency signalling link (e.g., 5G NR FR1):

\begin{enumerate}
    \item the BS commands each \textit{collaborative} MS entering the BS’s coverage area to use a certain analog beam $\mathbf{f}_{RF}$ (e.g., codebook index from $\mathbf{B}_{\mathrm{2D}}\left(N_1,N_2\right)$) and the relative training sequence. Additional information, such as position $\mathbf{p}_{\ell}$ and heading $\mathbf{\theta}_{\ell}$, could be requested to the MS by the BS;
    
    \item the \textit{collaborative} $\ell$-th MS, while moving in the BS's coverage area, transmits training sequences according to the BS using the fixed analog beam (see Figs. \ref{subfig:MVclustering1}, \ref{subfig:MVclustering2}, \ref{subfig:MVclustering3}). The position and heading, if requested, are related to the instant in which the training sequence is transmitted;
    
    \item the meantime the MS moves, the BS continuously scans all the analog beams $\mathbf{w}_{RF}$ of the Rx codebook. In the event of a match between BS and MS$_{\ell}$ analog beams, the BS stores: \textit{(i)} the received power $P_R$, \textit{(ii)} the MS$_{\ell}$ analog beam $\mathbf{f}_{RF}$, \textit{(iii)} the BS analog beam $\mathbf{w}_{RF}$, and \textit{(iv)} position $\mathbf{p}_{\ell}$ and heading $\mathbf{\theta}_{\ell}$;
    
    \item the BS, after the training period for multiple MSs, each with different precoder $\mathbf{f}_{RF}$, groups received powers in clusters based on positions and headings of the moving MSs and generates the related received power matrix $\mathbf{P}_R\left(\mathbf{\bar{p}}, \bar{\theta}\right)$, where $\mathbf{\bar{p}}$ and $\bar{\theta}$ are the reference position and heading respectively.  
    Fig. \ref{fig:ExamplePR} shows an example of $\mathbf{P}_R\left(\mathbf{\bar{p}}, \bar{\theta}\right)$ for FC-HBF and SC-HBF architectures.

\end{enumerate}

The optimal analog precoders/combiners $\mathbf{F}_{RF}$ and $\mathbf{W}_{RF}$ for each position are those maximizing the received power in matrix $\mathbf{P}_R\left(\mathbf{\bar{p}}, \bar{\theta}\right)$, learned from multiple passages illustrated in Figs. \ref{subfig:MVclustering1}, \ref{subfig:MVclustering2}, \ref{subfig:MVclustering3}. The problem consists in selecting the maximums of $\mathbf{P}_R\left(\mathbf{\bar{p}}, \bar{\theta}\right)$ corresponding to the true channel paths. In hybrid systems, however, as the full channel matrix $\mathbf{H}$ is unknown and cannot be directly estimated, it is not possible to approach the maximization analytically. Furthermore, heuristic approaches are disadvantageous, since $\mathbf{P}_R\left(\mathbf{\bar{p}}, \bar{\theta}\right)$ has several local maximums, due to LoS/NLoS spatial components of the channel (when they match with the selected beams) and their related grating lobes as can be observed in Fig.\ref{fig:ExamplePR}. 
Here, we select the set of beam pairs for $\mathbf{F}_{RF}$ and $\mathbf{W}_{RF}$ by searching for the first $N_T^{RF}$ and $N_R^{RF}$ maximums over the rows and the columns of $\mathbf{P}_R\left(\mathbf{\bar{p}}, \bar{\theta}\right)$ independently. This ensures that the analog beams at MS and BS are not repeated, i.e., we avoid rank-deficient precoders/combiners $\mathbf{F}_{RF}$/$\mathbf{W}_{RF}$ matrices.

Finally, the BS defines a list $\mathbf{L}_F$ of optimal analog precoders with the associated reference positions and headings, such that: 

\begin{equation}\label{eq:RFList}
    \left[\mathbf{L}_F\right]_k = \left\{\mathbf{F}_{RF,k}, \left(\mathbf{\bar{p}}_k, \bar{\theta}_k\right)\right\}
\end{equation}
and similarly for the optimal combiners, with list $\mathbf{L}_W$.

\section{Low-Rank Estimation of Digital Compressed Channels}\label{subsect:LR}

In the second stage of channel estimation, the BS has to learn the eigenmodes of the equivalent compressed channel $\widetilde{\mathbf{H}}$. Again, we exploit \textit{recurrent vehicle passages}. The BS sends the optimal analog precoders list $\mathbf{L}_{F}$ defined in the first stage to all collaborative MSs entering its coverage area. The $\ell$-th collaborative MS, while moving, transmits M consecutive training sequences $\mathbf{v}_{\ell}\in\mathbb{C}^{N_T^{RF}\times 1}$, such that $\widetilde{\mathbf{S}}_{\ell} = \left[\mathbf{v}_{1,\ell}, \mathbf{v}_{2,\ell}, \dots, \mathbf{v}_{M,\ell}\right] \in\mathbb{C}^{N_T^{RF}\times M}$. We assume the training sequences are chosen to be uncorrelated in both space and time and also mutually uncorrelated among different MSs, i.e., $\mathbb{E}[\widetilde{\mathbf{S}}_{\ell}\widetilde{\mathbf{S}}^{\mathrm{H}}_{s}] = \sigma^2_s\mathbf{I}_{N_T^{RF}}\delta_{s-\ell}$ and $\mathbb{E}[\widetilde{\mathbf{S}}^{\mathrm{H}}_{\ell}\widetilde{\mathbf{S}}_{s}] = \mathbf{I}_{M}\delta_{s-\ell}$, where $\sigma^2_s$ denotes the Tx power. The optimal analog precoder $\mathbf{F}_{RF}$ used for transmitting the training sequences is selected from the received list $\mathbf{L}_F$ based on the MS current MS position $\mathbf{p}_{\ell}$ and heading $\theta_{\ell}$.

The BS selects similarly the optimal analog combiner $\mathbf{W}_{RF}$ from the list $\mathbf{L}_W$ defined in the first stage, obtaining:
\begin{equation}\label{eq:receivedTrainingMat}
    \widetilde{\mathbf{Y}}_{\ell} = \widetilde{\mathbf{H}}_{\ell}\, \widetilde{\mathbf{S}}_{\ell} + \widetilde{\mathbf{N}}_{\ell}
\end{equation}
where the noise is such that $\mathbb{E}[\widetilde{\mathbf{N}}_{\ell}\widetilde{\mathbf{N}}^{\mathrm{H}}_{s}] = \widetilde{\mathbf{Q}}_n\delta_{s-\ell}$ and $\mathbb{E}[\widetilde{\mathbf{N}}^{\mathrm{H}}_{\ell}\widetilde{\mathbf{N}}_{s}] = \mathbf{I}_{M}\delta_{s-\ell}$.
At the end of the procedure, the set of received training sequences $\{\widetilde{\mathbf{Y}}_{\ell}\}_{\ell=1}^{\ell=L}$ for each position and heading is used to retrieve the compressed channel eigenmodes and the LR-estimated channel through algebraic manipulations, detailed in the following. In particular, two solutions are provided: \textit{(i)} optimal LR estimation, i.e., Joint Space (JS), and \textit{(ii)} sub-optimal LR estimation, i.e., Disjoint Space (DS). In Section \ref{sect:results}, the performance of the two approaches are compared and discussed. 

\subsection{Joint Space Low Rank (JS-LR) Estimation}\label{subsect:JSLR}

The LR-estimated compressed channel $\widehat{\mathbf{h}}_{\ell}\in\mathbb{C}^{N_R^{RF} N_T^{RF} \times 1}$ can be retrieved from the single received training signal $\widetilde{\mathbf{Y}}_{\ell}$ as the combination of a \textit{training sequence-dependent} matrix $\mathbf{G}_{\ell}$ and another one referred as \textit{position-dependent} matrix $\boldsymbol{\Pi}(\mathbf{\bar{p}},\bar{\theta})$, as \cite{Cerutti2020}:
\begin{equation}\label{eq:LR}
    \widehat{\mathbf{h}}_{\ell} = \boldsymbol{\Pi}(\mathbf{\bar{p}},\bar{\theta})\,\mathbf{G}_{\ell} \,\mathrm{vec}(\widetilde{\mathbf{Y}}_{\ell}) = \boldsymbol{\Pi}(\mathbf{\bar{p}},\bar{\theta}) \, \overline{\mathbf{y}}_{\ell} ,
\end{equation}
where $\overline{\mathbf{y}}_{\ell}\in \mathbb{C}^{N_R^{RF} N_T^{RF} \times 1}$ is the pre-processed sequence by $\mathbf{G}_{\ell}$. A notable example is the LS or U-ML channel estimation, where $\mathbf{G}_{\ell}$ is a suitable rearrangement of known pilot symbols. For instance, for the LS channel estimation from \eqref{eq:receivedTrainingMat}, $\mathbf{G}_{\ell} = (\widetilde{\mathbf{S}}^{\mathrm{T}}_{\ell} \otimes \mathbf{I}_{N_R^{RF}} )^{\dagger}$. 

The position-dependent linear processing $\boldsymbol{\Pi}(\mathbf{\bar{p}},\bar{\theta})$ is estimated from an ensemble of $L$ training sequences $\{\overline{\mathbf{y}}_{\ell}\}_{\ell=1}^{\ell=L}$, originated from multiple vehicles passing in the same location such that each one has the same propagation structure with all the others. In the context of \eqref{eq:LR}, matrix $\boldsymbol{\Pi}(\mathbf{\bar{p}},\bar{\theta})$ operates a modal filtering on $\overline{\mathbf{y}}_{\ell}$, projecting it onto the propagation subspace \cite{Nicoli2003}.

The first step to obtain the position-dependent $\boldsymbol{\Pi}(\mathbf{\bar{p}},\bar{\theta})$ is to identify the algebraic structure of the compressed channel $\widetilde{\mathbf{h}} = \mathrm{vec}(\widetilde{\mathbf{H}})$, which can be shown to be:
\begin{equation}\label{eq:compressed_channel_vector}
\begin{split}
    \widetilde{\mathbf{h}} & = \left(\mathbf{F}^{\mathrm{T}}_{RF}\otimes \mathbf{W}^{\mathrm{H}}_{RF}\right)\boldsymbol{\mathcal{A}}\left(\boldsymbol{\psi},\boldsymbol{\vartheta}\right)\boldsymbol{\alpha} = \mathbf{T}\left(\boldsymbol{\psi},\boldsymbol{\vartheta}\right) \,\boldsymbol{\alpha}
\end{split}
\end{equation}
where \textit{(i)} $\mathbf{T}\left(\boldsymbol{\psi},\boldsymbol{\vartheta}\right)\in \mathbb{C}^{N_T^{RF} N_R^{RF} \times P}$ embeds the spatial features of the compressed channel, invariant across multiple MSs passing the same position; \textit{(ii)} matrix $\boldsymbol{\mathcal{A}}\left(\boldsymbol{\psi},\boldsymbol{\vartheta}\right)\in\mathbb{C}^{N_T N_R \times P} = \mathbf{A}_T\left(\boldsymbol{\psi}\right) \diamond \mathbf{A}_R\left(\boldsymbol{\vartheta}\right)$, and \textit{(iii)} $\boldsymbol{\alpha} \in\mathbb{C}^{P\times 1}= \left[\alpha_1,\dots,\alpha_P\right]^{\mathrm{T}}$ collects the channel amplitudes, different from MS to MS but with the same power profile \eqref{eq:scatteringamplitude_corr}.

Let us define the compressed channel correlation $\widetilde{\mathbf{R}} = \mathbb{E}[\widetilde{\mathbf{h}} \, \widetilde{\mathbf{h}}^{\mathrm{H}}]$, which can be computed by exploiting the invariance of AoAs/AoDs across multiple vehicles, as:
\begin{equation}\label{eq:SF_compressed_channel_vector_correlation_true}
\begin{split}
    \widetilde{\mathbf{R}} &= \mathbf{T}\left(\boldsymbol{\psi},\boldsymbol{\vartheta}\right) \mathbf{P}\, \mathbf{T}\left(\boldsymbol{\psi},\boldsymbol{\vartheta}\right)^{\mathrm{H}} = \\ &= \sum_{p=1}^P P_p \left[\mathbf{F}^{\mathrm{T}}_{RF}\mathbf{R}_{T,p}\mathbf{F}^{*}_{RF} \otimes \mathbf{W}^{\mathrm{H}}_{RF}\mathbf{R}_{R,p}\mathbf{W}_{RF}\right]
\end{split}
\end{equation}
where $\mathbf{R}_{T,p} \in \mathbb{C}^{N_T \times N_T} = \mathbf{a}_T\left(\boldsymbol{\psi}_p\right)\mathbf{a}_T\left(\boldsymbol{\psi}_p\right)^{\mathrm{H}}$ and $\mathbf{R}_{R,p} \in \mathbb{C}^{N_R \times N_R} = \mathbf{a}_R\left(\boldsymbol{\vartheta}_p\right)\mathbf{a}_R\left(\boldsymbol{\vartheta}_p\right)^{\mathrm{H}}$. 

We can re-parameterize the channel $\widetilde{\mathbf{h}}$ in \eqref{eq:compressed_channel_vector} using the $\widetilde{r}$ leading eigenvectors of $\widetilde{\mathbf{R}}$, i.e., $\widetilde{\mathbf{U}} = \mathrm{eig}_{\widetilde{r}}(\widetilde{\mathbf{R}})$, such that:
\begin{equation}\label{eq:JS_subspace}
\begin{split}
    &\mathrm{span}(\widetilde{\mathbf{U}}) = \mathrm{span}(\mathbf{T}\left(\boldsymbol{\psi},\boldsymbol{\vartheta}\right)),
\end{split}
\end{equation}
the orthonormal basis $\widetilde{\mathbf{U}}\in\mathbb{C}^{N_T^{RF} N_R^{RF} \times \widetilde{r}}$ span the \textit{joint} Tx \textit{and} Rx subspace of the compressed channel, of dimension $\widetilde{r} = \mathrm{rank}(\widetilde{\mathbf{R}}) = \mathrm{rank}(\mathbf{T}\left(\boldsymbol{\psi},\boldsymbol{\vartheta}\right))$. The latter represents the number of compressed channel paths (diversity order) that can be resolved by the digital system:
\begin{equation}\label{eq:Req_rank}
    \widetilde{r} \leq \mathrm{min}\left(\mathrm{rank}(\mathbf{F}_{RF}) \mathrm{rank}(\mathbf{W}_{RF}), r \right),
\end{equation}
where $r =  \mathrm{rank}(\boldsymbol{\mathcal{A}}\left(\boldsymbol{\psi},\boldsymbol{\vartheta}\right)) \leq \mathrm{min}(N_T N_R , P)$ is the number of resolvable paths of the full channel $\mathbf{h} = \boldsymbol{\mathcal{A}}\left(\boldsymbol{\psi},\boldsymbol{\vartheta}\right)\boldsymbol{\alpha}$, obtained by rearranging \eqref{eq:channel_matrix} similarly to \eqref{eq:compressed_channel_vector}.

From the LR contraint \eqref{eq:Req_rank}, the position-dependent matrix $\boldsymbol{\Pi}(\mathbf{\bar{p}},\bar{\theta})$ is estimated as \cite{Cerutti2020}:
\begin{equation}\label{eq:JS_projector}
    \widehat{\boldsymbol{\Pi}}(\mathbf{\bar{p}},\bar{\theta}) = \mathbf{C}^{\frac{\mathrm{H}}{2}} \, \widehat{\boldsymbol{\Pi}}_{\mathrm{JS}} \, \mathbf{C}^{-\frac{\mathrm{H}}{2}},
\end{equation}
where: 
\begin{itemize}

    \item $\mathbf{C}$ is the covariance matrix of $\overline{\mathbf{y}}_{\ell}$, corresponding to the Cramer-Rao Bound (CRB). For this problem, it is $\mathbf{C} \approx (\mathbf{I}_{N_T^{RF}}/\sigma_s^2)  \otimes \widetilde{\mathbf{Q}}_n$ (asymptotic approximation). Matrix $\mathbf{C}$ is used to perform the whitening (and de-whitening) of $\overline{\mathbf{y}}_{\ell}$ to optimally handle any presence of noise correlation (e.g., interference);
    
    \item $\widehat{\boldsymbol{\Pi}}_{\mathrm{JS}} = \widehat{\mathbf{U}}\widehat{\mathbf{U}}^{\mathrm{H}}$ is the JS-LR projection matrix onto the propagation subspace spanned by $\widehat{\mathbf{U}}= \mathrm{eig}_{\widetilde{r}}(\widehat{\mathbf{R}})$, where 
    \begin{equation}\label{eq:STchannel_vector_correlation_sample}
        \widehat{\mathbf{R}} =  \frac{1}{L}\sum_{\ell=1}^{L}
         \overline{\overline{\mathbf{y}}}_{\ell} \overline{\overline{\mathbf{y}}}_{\ell}^{\mathrm{H}}
    \end{equation}
    is the sample correlation of \textit{whitened} sequences $\overline{\overline{\mathbf{y}}}_{\ell} = \mathbf{C}^{-\frac{\mathrm{H}}{2}} \overline{\mathbf{y}}_{\ell}$, collected from $L$ different MSs passing the same position. 
    
\end{itemize}

The performance of the proposed LR channel estimation, hereafter referred to as Joint-Space LR (JS-LR), provided by the application of $\boldsymbol{\Pi}(\mathbf{\bar{p}},\bar{\theta})$ in \eqref{eq:JS_projector} on signal $\overline{\mathbf{y}}_{\ell}$, depends on the \textit{sparsity degree} of the compressed channel $\widetilde{\mathbf{h}}$. 
The latter is proportional to the ratio between the effective number of \textit{spatially-separable} analog beams for MS and BS, respectively $N_T^{\mathrm{beams}}$ and $N_R^{\mathrm{beams}}$, and the number of resolvable paths $\widetilde{r}$ of the compressed channel.

For FC-HBF systems, the number of separable beams are $N_T^{\mathrm{beams}}\leq \mathrm{rank}(\mathbf{F}_{RF})\leq N_T^{RF}$, $N_R^{\mathrm{beams}}\leq\mathrm{rank}(\mathbf{W}_{RF})\leq N_R^{RF}$ as the Tx/Rx terminals can, in general, use arbitrary angular separated analog beams. For the analog beams chosen here as selected from orthogonal codebooks and not repeated (Section \ref{subsect:beamalignment}), we have 
$N_T^{\mathrm{beams}} = \mathrm{rank}(\mathbf{F}_{RF}) = N_T^{RF}$ and $N_R^{\mathrm{beams}} = \mathrm{rank}(\mathbf{W}_{RF}) = N_R^{RF}$, and the sparsity degree of the compressed channel is maximum. 

In SC-HBF architectures, the block-diagonal structure of $\mathbf{F}_{RF}$ and $\mathbf{W}_{RF}$ leads, in general, to $N^{\mathrm{beams}}_T \leq \mathrm{rank}(\mathbf{F}_{RF}) = N_T^{RF}$ and $N^{\mathrm{beams}}_R \leq \mathrm{rank}(\mathbf{F}_{RF}) = N_R^{RF}$, but again the proposed analog beam alignment ensures that $N^{\mathrm{beams}}_T = \mathrm{rank}(\mathbf{F}_{RF}) = N_T^{RF}$ and $N^{\mathrm{beams}}_R = \mathrm{rank}(\mathbf{F}_{RF}) = N_R^{RF}$, as every RF chain employs a different orthogonal beam, maximizing the channel sparsity. 

In this regard, provided that:
\begin{equation}\label{eq:JS_sparsity}
    \widetilde{r} < N_T^{RF} N_R^{RF},
\end{equation}
the LR provides superior performance compared to conventional approaches (e.g., LS/U-ML).

\subsection{Disjoint Space Low Rank Estimation}\label{subsect:DSLR}

To reduce the complexity of the JS-LR implementation, mainly due to the large-matrix eigendecomposition of $\widehat{\mathbf{R}}$ in \eqref{eq:STchannel_vector_correlation_sample}, we propose a sub-optimal approach, referred herein as Disjoint-Space LR (DS-LR). This assumes the separability of Tx and Rx spatial subspaces of the compressed channel as suggested in \cite{Nicoli2003}. In analogy to \eqref{eq:SF_compressed_channel_vector_correlation_true}, we leverage the algebraic structure of $\widetilde{\mathbf{H}}$ in \eqref{eq:compressed_channel_matrix} and define the Tx and Rx compressed channel correlations $\widetilde{\mathbf{R}}_{T} = \mathbb{E}[\widetilde{\mathbf{H}}^{\mathrm{H}} \widetilde{\mathbf{H}}]$ and $\widetilde{\mathbf{R}}_{R} = \mathbb{E}[\widetilde{\mathbf{H}} \widetilde{\mathbf{H}}^{\mathrm{H}}]$, respectively equal to:
\begin{align}
\begin{split}
\widetilde{\mathbf{R}}_{T}
& = \mathbf{F}_{RF}^{\mathrm{H}} \mathbf{A}^*_T\left(\boldsymbol{\psi}\right) \, \widetilde{\boldsymbol{\Gamma}}_{T} \, \mathbf{A}^{\mathrm{T}}_T\left(\boldsymbol{\psi}\right) \mathbf{F}_{RF},
\end{split} \label{eq:S_Tx_compressed_channel_correlation_true} \\
\begin{split}
\widetilde{\mathbf{R}}_{R} 
& = \mathbf{W}_{RF}^{\mathrm{H}} \mathbf{A}_R\left(\boldsymbol{\vartheta}\right) \, \widetilde{\boldsymbol{\Gamma}}_{R} \, \mathbf{A}^{\mathrm{H}}_R\left(\boldsymbol{\vartheta}\right) \mathbf{W}_{RF},
\end{split}\label{eq:S_Rx_compressed_channel_correlation_true} 
\end{align}
where
\begin{align}
\widetilde{\boldsymbol{\Gamma}}_{T} & = \mathbf{P} \odot \mathbf{A}^{\mathrm{H}}_R\left(\boldsymbol{\vartheta}\right)\mathbf{W}_{RF}\mathbf{W}_{RF}^{\mathrm{H}}\mathbf{A}_R\left(\boldsymbol{\vartheta}\right),\label{eq:S_Tx_compressed_channel_correlation_true_eigval}\\
\widetilde{\boldsymbol{\Gamma}}_{R} & = \mathbf{P} \odot \mathbf{A}^{\mathrm{T}}_T\left(\boldsymbol{\psi}\right)\mathbf{F}_{RF}\mathbf{F}_{RF}^{\mathrm{H}}\mathbf{A}^{*}_T\left(\boldsymbol{\psi}\right),\label{eq:S_Rx_compressed_channel_correlation_true_eigval}
\end{align}
are diagonal matrices of $P\times P$ size. Eq. \eqref{eq:S_Tx_compressed_channel_correlation_true_eigval}-\eqref{eq:S_Rx_compressed_channel_correlation_true_eigval} highlight how the analog precoder/combiner pair affects the eigenvalues of the Tx and Rx channel correlation matrices. The last term in \eqref{eq:S_Tx_compressed_channel_correlation_true_eigval}, for instance, represents the overall matching between the steering vectors of the AoAs with the combiner $\mathbf{W}_{RF}$: for a fixed precoder $\mathbf{F}_{RF}$, the Tx side experiences different channel gains (sum of eigenvalues) for different combiners. The same applies for the Rx side, with an optimum precoder/combiner pair for the best pointing between Tx and Rx.

From the $\widetilde{r}_T$ and $\widetilde{r}_R$ leading eigenvectors of $\widetilde{\mathbf{R}}_{T}$ and $\widetilde{\mathbf{R}}_{R}$, i.e., $\widetilde{\mathbf{U}}_T = \mathrm{eig}_{\widetilde{r}_T}(\widetilde{\mathbf{R}}_{T})$ and $\widetilde{\mathbf{U}}_R = \mathrm{eig}_{\widetilde{r}_R}(\widetilde{\mathbf{R}}_{R})$, we have that:
\begin{align}
    \mathrm{span}(\widetilde{\mathbf{U}}_T)&  = \mathrm{span}\left(\mathbf{A}_{T}^{\mathrm{T}}(\boldsymbol{\psi}) \mathbf{F}_{RF}\right),\\
    \mathrm{span}(\widetilde{\mathbf{U}}_R)&  = \mathrm{span}\left(\mathbf{W}_{RF}^{\mathrm{H}} \mathbf{A}_R(\boldsymbol{\vartheta})\right),
\end{align}
i.e., $\widetilde{\mathbf{U}}_T\in\mathbb{C}^{N_T^{RF} \times \widetilde{r}_T}$ and $\widetilde{\mathbf{U}}_R\in\mathbb{C}^{N_R^{RF} \times \widetilde{r}_R}$ span the Tx and Rx spatial subspaces of the compressed channel $\widetilde{\mathbf{H}}$, of dimensions $\widetilde{r}_T$ and $\widetilde{r}_R$ (see \eqref{eq:Req_rank_DS_Tx}-\eqref{eq:Req_rank_DS_Rx} in Section \ref{sect:system_model}).

The position-dependent matrix for the DS-LR method is:
\begin{equation}\label{eq:DS_projector}
    \widehat{\boldsymbol{\Pi}}(\mathbf{\bar{p}},\bar{\theta}) = \mathbf{C}^{\frac{\mathrm{H}}{2}}\,  \widehat{\boldsymbol{\Pi}}_{\mathrm{DS}} \, \mathbf{C}^{-\frac{\mathrm{H}}{2}},
\end{equation}
where $\widehat{\boldsymbol{\Pi}}_{\mathrm{DS}} = \widehat{\mathbf{U}}^*_{T}\widehat{\mathbf{U}}^{\mathrm{T}}_{T}\otimes \widehat{\mathbf{U}}_{R}\widehat{\mathbf{U}}^{\mathrm{H}}_{R}$ is the DS-LR projector onto to the propagation subspace, represented by basis $\widehat{\mathbf{U}}^*_{T}\otimes \widehat{\mathbf{U}}_{R}$. Notice that the Kronecker separability of Tx and Rx subspaces is an approximation, as the Kronecker structure of the digital channel correlation \eqref{eq:SF_compressed_channel_vector_correlation_true} holds for single paths only. Similarly to JS-LR, $\widehat{\mathbf{U}}_{T} = \mathrm{eig}_{\widetilde{r}_{T}}(\widehat{\mathbf{R}}_{T}) $ and $\widehat{\mathbf{U}}_{R} = \mathrm{eig}_{\widetilde{r}_{R}}(\widehat{\mathbf{R}}_{R})$ are set from the $\widetilde{r}_{T}$ and $\widetilde{r}_{R}$ leading eigenvectors of the following sample correlations: 
\begin{align}
 \widehat{\mathbf{R}}_{T} &= \frac{1}{L}\sum_{\ell=1}^{L}  \overline{\overline{\mathbf{Y}}}^{\mathrm{H}}_{\ell}\, \overline{\overline{\mathbf{Y}}}_{\ell}, \label{eq:SST_compressed_channel_correlations_sample_1}\\
 \widehat{\mathbf{R}}_{R} &= \frac{1}{L}\sum_{\ell=1}^{L}  \overline{\overline{\mathbf{Y}}}_{\ell} \, \overline{\overline{\mathbf{Y}}}^{\mathrm{H}}_{\ell}, \label{eq:SST_compressed_channel_correlations_sample_2}
\end{align}
where we indicate with $\overline{\overline{\mathbf{Y}}}_{\ell}\in\mathbb{C}^{N_R^{RF}\times N_T^{RF}} = \mathrm{vec}^{-1}(\overline{\overline{\mathbf{y}}}_{\ell})$ the whitened sequences in matrix form. It can be demonstrated that, asymptotically ($L\rightarrow \infty$):
\begin{align}
    &\mathrm{span}(\widehat{\mathbf{U}}_T) \rightarrow \mathrm{span}(\mathbf{A}_{T}^{\mathrm{T}}(\boldsymbol{\psi}) \mathbf{F}_{RF}),\\
    &\mathrm{span}(\widehat{\mathbf{U}}_R) \rightarrow \mathrm{span}(\widetilde{\mathbf{Q}}^{-\frac{\mathrm{H}}{2}}_n\mathbf{W}_{RF}^{\mathrm{H}} \mathbf{A}_R(\boldsymbol{\vartheta})).
\end{align}

The DS-LR channel estimation method provides a performance gain with respect to conventional approaches when the spatial structure of $\widetilde{\mathbf{H}}$ is sparse, which is equivalent to state that at least one of the following conditions hold:
\begin{align}
    \widetilde{r}_{T} & < N_T^{RF}\label{eq:DS_sparsity_1},\\
    \widetilde{r}_{R} & < N_R^{RF}\label{eq:DS_sparsity_2}.
\end{align}
Compared to JS-LR, the DS-LR method requires a lower number of training sequences, $L$, to estimate the compressed channel eigenmodes, at the price of a reduced performance (the sparsity degree of DS-LR is always less than the JS-LR one). 

\textit{Remark}. Without interference ($\mathbf{Q}_n = \sigma_n^2 \mathbf{I}_{N_R}$) and with an orthogonal codebook for $\mathbf{W}_{RF}$ as here, it follows that $\widetilde{\mathbf{Q}}_n \approx N \sigma_n^2 \mathbf{I}_{N^{RF}_R}$, where $\sigma_n^2$ is the noise power and $N$ accounts for analog beamforming ($N=N_R$ for FC-HBF, $N=N^B_R$ for SC-HBF). In this setting, for $L \rightarrow \infty$ we have:
\begin{align}
    &\mathrm{span}(\widehat{\mathbf{U}}) \rightarrow \mathrm{span}(\widetilde{\mathbf{U}}),\\
    &\mathrm{span}(\widehat{\mathbf{U}}_{T}) \rightarrow \mathrm{span}(\widetilde{\mathbf{U}}_T),\\
    &\mathrm{span}(\widehat{\mathbf{U}}_{R}) \rightarrow \mathrm{span}(\widetilde{\mathbf{U}}_R).
\end{align}
Therefore, the whitening/de-whitening reduces the position-dependent matrix $\boldsymbol{\Pi}(\mathbf{\bar{p}},\bar{\theta})$ to the projection matrix associated to the sample estimates of \eqref{eq:SF_compressed_channel_vector_correlation_true} (JS-LR) and \eqref{eq:S_Tx_compressed_channel_correlation_true}-\eqref{eq:S_Rx_compressed_channel_correlation_true} (DS-LR). 

\subsection{Lossy vs. Lossless Channel Compression for FC-HBF Architectures}\label{subsect:lossylossless}
By exploiting \eqref{eq:Req_rank_DS_Tx}, \eqref{eq:Req_rank_DS_Rx} and \eqref{eq:analog_precodercombiner_rank} we can observe that if both the following conditions hold

\begin{align}
   & N_T^{RF} \geq r_T, \label{eq:hp1} \\
   & N_R^{RF} \geq r_R, \label{eq:hp2}
\end{align}\label{eq:hp}

this implies that there exist an analog precoder/combiner $\mathbf{F}_{RF}/\mathbf{W}_{RF}$ with $\mathrm{rank}(\mathbf{F}_{RF}) \geq r_T$ and $\mathrm{rank}(\mathbf{W}_{RF}) \geq r_R$, such that FC-HBF performance, in terms of Spectral Efficiency (SE), attains the Full-Digital (FD) one. The first condition, \eqref{eq:hp1}, asserts that the overall number of RF chains at Tx must be larger than the number of Tx-resolvable paths of the \textit{full} channel $\mathbf{H}$. This is derived from \eqref{eq:Req_rank_DS_Tx} by noticing that, if \eqref{eq:hp1} does not hold, $\widetilde{r}_{T} < r_T$ would mean that the HBF system cannot explore the full channel diversity for insufficient number of available beams at Tx, regardless the choice of $\mathbf{F}_{RF}/\mathbf{W}_{RF}$. This is equivalent to a \textit{lossy} compression of the channel. Condition \eqref{eq:hp2} can be analogously derived from \eqref{eq:Req_rank_DS_Rx}. When \textit{both} \eqref{eq:hp1} and \eqref{eq:hp2} apply, $\widetilde{r}_{T} \leq r_T$ and $\widetilde{r}_{R} \leq r_R$, i.e., the number of resolvable paths at Tx and Rx before and after the analog compression can be equal when a suitable combination of $\mathbf{F}_{RF}/\mathbf{W}_{RF}$ is employed (\textit{lossless} compression of the channel). 

\textit{Remark 1}: In practical FC-HBF systems, where the analog precoder/combiner are defined by a codebook, the performances can deteriorate if the resolution is poor, i.e., low angular sampling interval.

\textit{Remark 2}: The previous consideration does not apply to SC-HBF systems, unless a proper Tx power augmentation is considered. Indeed, for a fixed Tx power, the reduced beamforming gain of SC-HBF compared to FC-HBF does not allow to reach the performance of FD systems.

\subsection{Digital Precoders/Combiners Design}\label{subsect:digitalPrec&Comb}

The digital precoders/combiners are retrieved from the LR estimated compressed channel matrix $\widehat{\mathbf{H}} = \mathrm{vec}^{-1}(\widehat{\mathbf{h}})$, which must be known at the Tx side through a feedback from Rx. The optimal digital precoder $\mathbf{F}_{BB}$ is \cite{1658244}:
\begin{equation}\label{eq:DigitalPrecoderDesign}
    \mathbf{F}_{BB} = \mathrm{eig}_{N_S}\left(\widehat{\mathbf{H}}^{\mathrm{H}} \widehat{\mathbf{H}}\right),
\end{equation}
while the digital combiner $\mathbf{W}_{BB}^{\mathrm{H}}$ is derived using the optimal unconstrained Minimum Mean Squared Error (MMSE) as \cite{BigS}
\begin{equation}\label{eq:DigitalCombinerDesign}
\begin{split}
  \mathbf{W}_{BB}^{\mathrm{H}} = \bigg( \widehat{\mathbf{H}}^{\mathrm{H}} \mathbf{F}^{\mathrm{H}}_{BB} \widetilde{\mathbf{Q}}_n^{-1} \mathbf{F}_{BB}  & \widehat{\mathbf{H}} + \frac{\mathbf{I}_{N_S}}{N_S} \bigg)^{-1} \widehat{\mathbf{H}}^{\mathrm{H}} \mathbf{F}^{\mathrm{H}}_{BB}\widetilde{\mathbf{Q}}_n^{-1}.
\end{split}
\end{equation}

\section{Numerical Results}\label{sect:results}

\begin{figure}[!b]
    \centering
    \includegraphics[width=0.5\columnwidth]{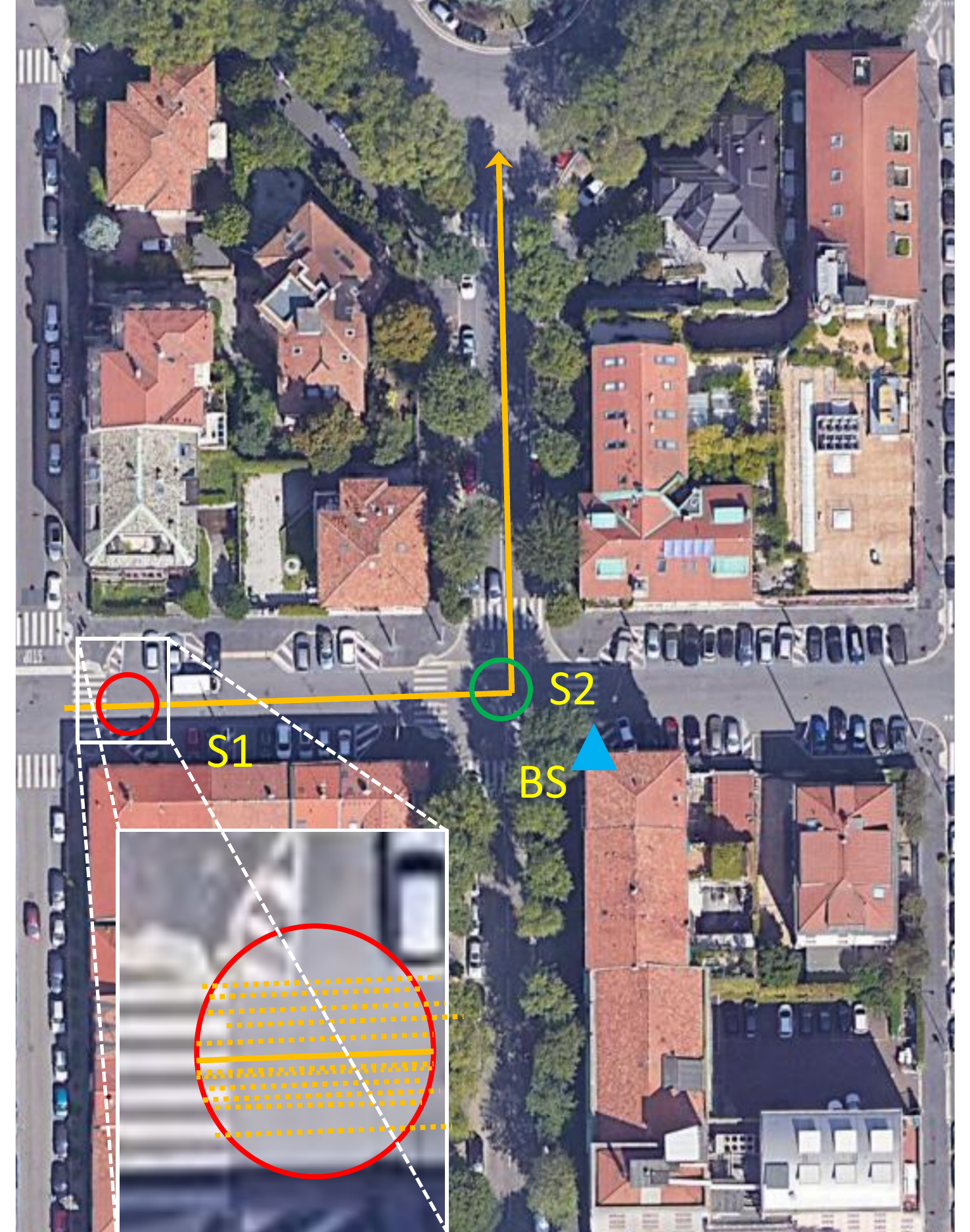}
    \caption{Urban scenario used in simulations: the solid yellow line represents the reference vehicle trajectory, the blue triangle the mmW BS (located at 6 m height from ground) while red and green circles the S1 and S2 locations used for testing the proposed LR channel estimation. The inset sketches the trajectories dataset used to simulate multiple vehicle passages in the ray-tracing software }
    \label{fig:street}
\end{figure}

\begin{figure}[!t]
    \centering
    \includegraphics[width=0.48\columnwidth]{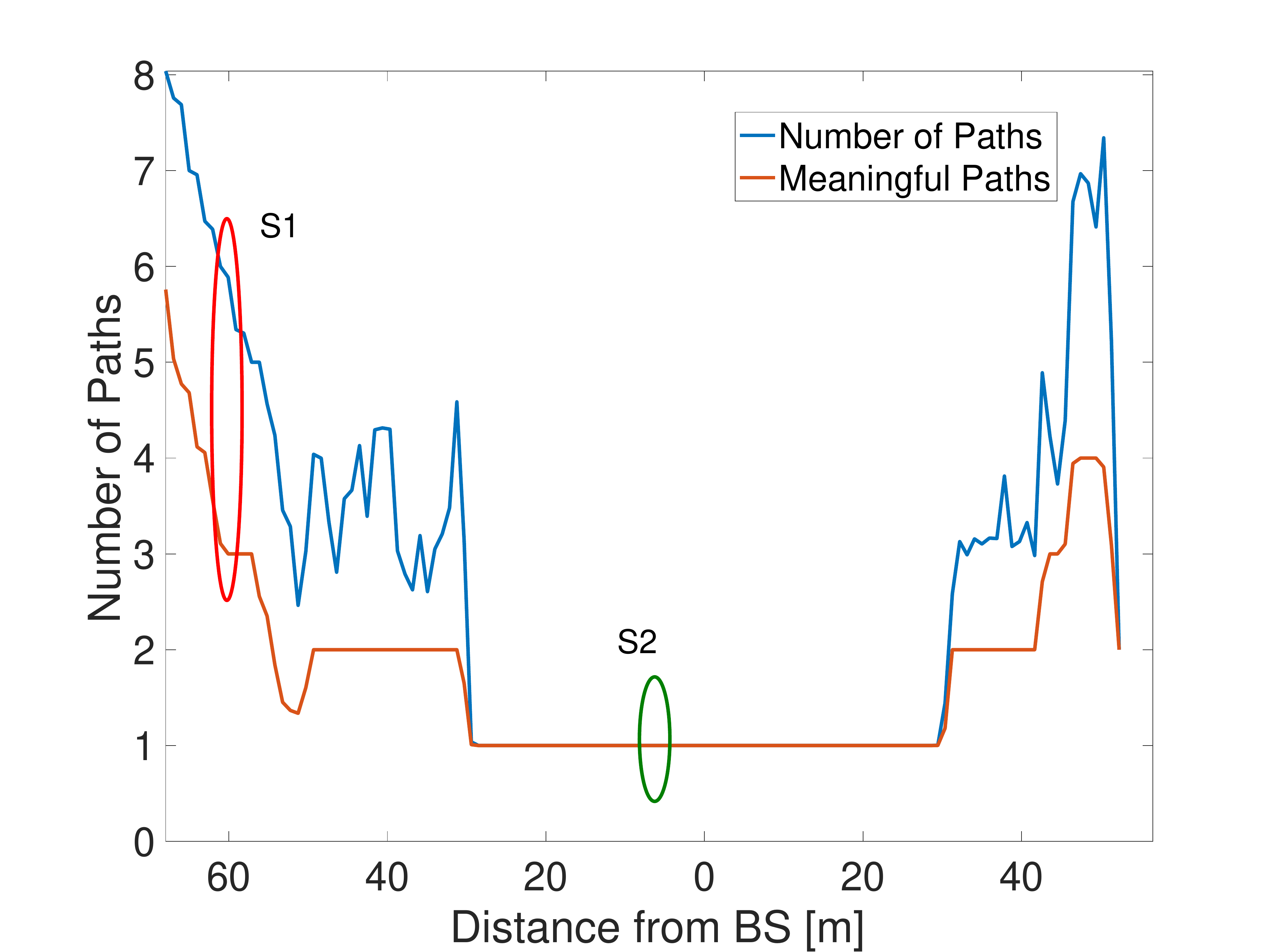}
    \caption{Number of channel paths vs. distance from the BS: the blue curve represent the number of paths provided by ray-tracer, the red one the meaningful ones}
    \label{fig:paths}
\end{figure}

\begin{figure}[b!]
   \centering
   \subfloat[]{\includegraphics[width=.3\columnwidth]{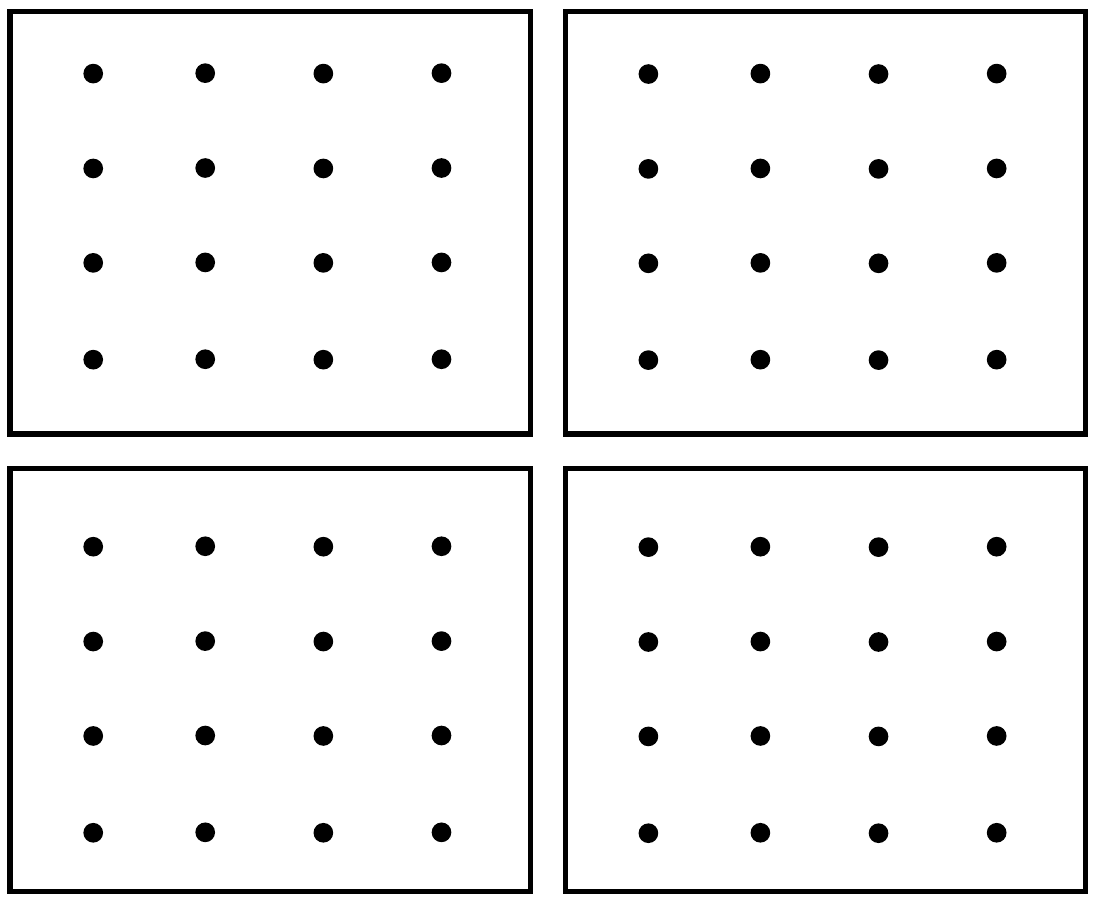}\label{subfig:Txconfig}}\hspace{0.2mm}
   \subfloat[]{\includegraphics[width=.3 \columnwidth]{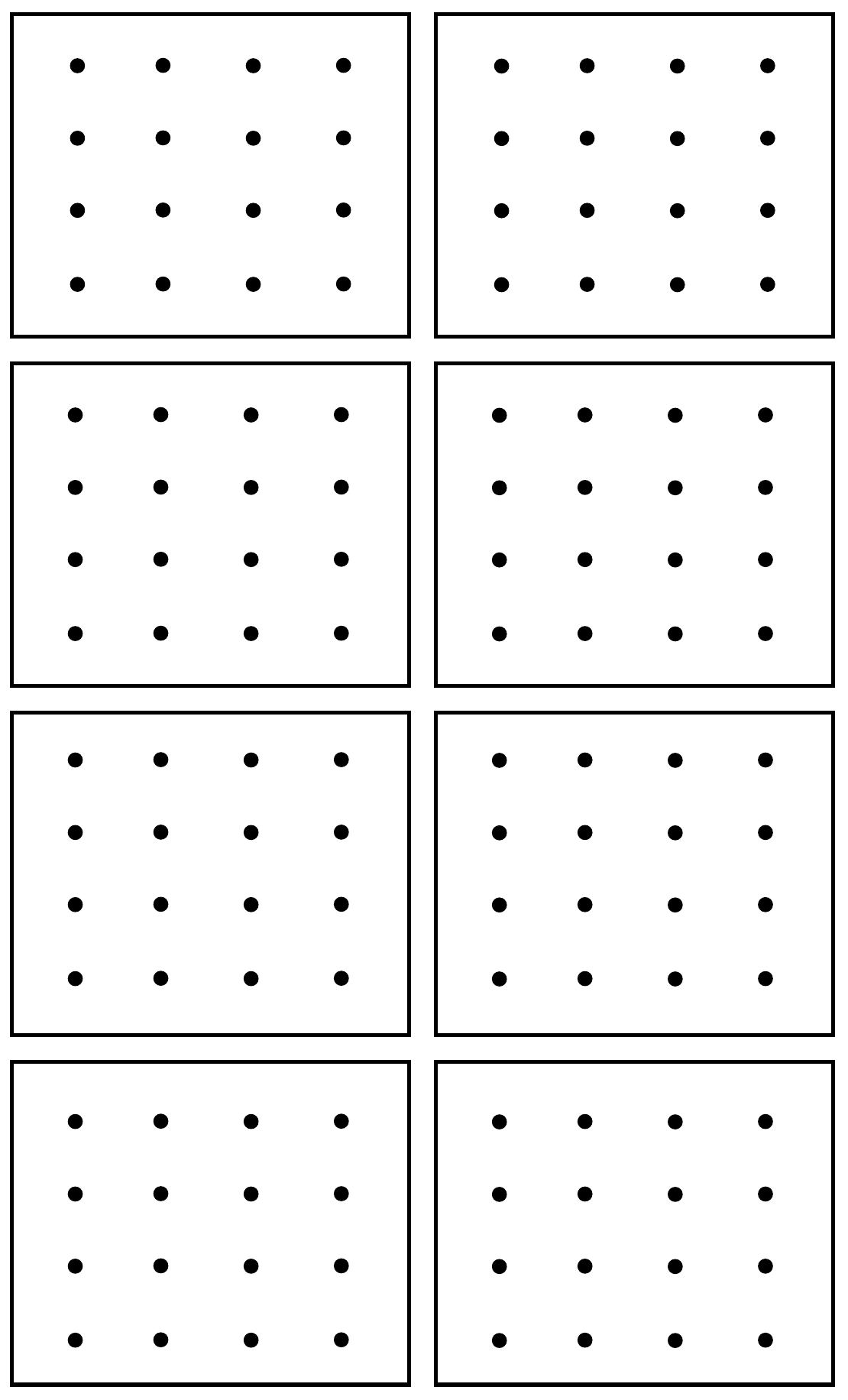}\label{subfig:Rxconfig}}    
   \caption{Hybrid arrays configuration for MS (\ref{subfig:Txconfig}) and BS (\ref{subfig:Rxconfig})}    \label{fig:TXRXConfiguration}
\end{figure}

To demonstrate the effectiveness of the proposed channel estimation method, we present the results obtained through numerical simulations using ray-tracing channel data and a set of realistic vehicle trajectories. The latter are aimed at simulating multiple vehicle passages, in a typical urban scenario (Fig. \ref{fig:street}). Two locations are selected for testing: the first (red circle in Fig. \ref{fig:street}) relatively far from the BS, $\approx 60$ m, and the second (green circle) at $\approx 8$ m (close to the BS). To ease the reader in analyzing the results, we will refer to these locations as S1 (far from the BS) and S2 (close to the BS). Fig. \ref{fig:paths} shows the number of channel paths as function of the MS-to-BS distance. The blue curve represents the number of paths provided by the ray-tracer, while the red curve is the number of meaningful paths, i.e., those with cumulative power within the $99.9$ percentile. Red and green ellipses in Fig. \ref{fig:paths} identify the S1 and S2 locations used in simulation.

We consider an interference-free MS-to-BS (UL) communication in the 5G NR FR2 band (28 GHz carrier frequency) \cite{TS38213}. The hybrid MIMO setting is such that the BS is equipped with $N_R=128$ ($16 \times 8$) antennas and $N_R^{RF}=8$ RF chains, while the MS (i.e., each vehicle) with $N_T=64$ ($8 \times 8$) antennas and $N_T^{RF}=4$ RF chains. We make use of Altair WinProp ray-tracing software \cite{WinProp} to generate the required channel data (power, AoDs, AoAs and scattering amplitude for each ray), whereas the MIMO channel \eqref{eq:channel_matrix} is obtained by post-processing in far-field (i.e., AoDs/AoAs equal for each Tx/Rx antennas). The required trajectories (i.e., position, velocity and direction over time) are instead generated by means of SUMO \cite{SUMO2018}. In both MV beam alignment (Section \ref{subsect:beamalignment}) and LR training (Section \ref{subsect:LR}) procedures, we consider the assignment of MSs' positions according to the spatial granularity of the experiment. In other words, we exploit multiple vehicle passages in a spatial region of radius $\rho$, where AoAs/AoDs are invariant. We set $\rho=2$ m for S1 and $\rho=0.5$ m for S2, determined with ray-tracing. The size of the MV region plays an important role in the proposed method: if excessive, a performance penalty is experienced by the system as the channel subspaces decorrelate (Section \ref{subsect:S2}).

Unless otherwise specified, the parameters given in Table \ref{tab:SimParam} are used to generate the results, while the MS and BS array configurations are in Fig. \ref{fig:TXRXConfiguration}. Most of the results we present in this section are related to the multipath scenario S1, while the single-path S2 is used for comparison.

\begin{table}[t!]
    \centering
    \caption{Simulation Parameters}
    \begin{tabular}{l|c|c}
    \toprule
        \textbf{Parameter} &  \textbf{Symbol} & \textbf{Value(s)}\\
        \hline
        Carrier frequency & $f_c$  & $28$ GHz \\
        Number of channel paths &$P$ & $1-7$\\
        Number of data streams & $N_S$ & 1 \\
        Number of MS antennas & $N_T$ & $64$ ($8 \times 8$)\\
        Number of MS RF chains & $N_T^{RF}$ & $4$\\
        Number of MS antennas x sub-array & $N_T^{B}$ & $16$\\
        Number of BS antennas & $N_R$ & $128$ ($16 \times 8$)\\
        Number of BS RF chains & $N_R^{RF}$ & $8$\\
        Number of BS antennas x sub-array & $N_R^{B}$ & $16$\\
        BS height from ground & - & 6 m\\
        MV region radius & $\rho$ & 2 m (S1), 0.5 m (S2)\\
        \bottomrule
    \end{tabular}
    \label{tab:SimParam}
\end{table}

The performance of both JS-LR and DS-LR channel estimation methods are compared to the U-ML one in terms of Spectral Efficiency (SE) and MSE on compressed channel estimation. The SE is defined as \cite{1203154}:

\begin{equation}\label{eq:SE}
    \begin{split}
        R = \log_2\bigg|\mathbf{I}_{N_S} + \frac{1}{N_S} \mathbf{Q}^{-1}_{\mathrm{eff}} \mathbf{H}_{\mathrm{eff}} \mathbf{H}_{\mathrm{eff}}^{\mathrm{H}}\bigg|
    \end{split}
\end{equation}

where $\mathbf{Q}_{\mathrm{eff}} = \mathbf{W}^{\mathrm{H}}_{BB}\widetilde{\mathbf{Q}}_{n} \mathbf{W}_{BB}$ is the covariance of the noise at the decision variable, and $\mathbf{H}_{\mathrm{eff}} = \mathbf{W}_{BB}^{\mathrm{H}} \widetilde{\mathbf{H}} \mathbf{F}_{BB}$ is the effective end-to-end channel between MS and BS. The MSE is computed as:

\begin{equation}\label{eq:MSE}
    \mathrm{MSE} =  \mathbb{E}\left[\norm{\widehat{\mathbf{h}}- \widetilde{\mathbf{h}}}^2\right],
\end{equation}

and for U-ML method it is compared to the theoretical Cramer-Rao Lower Bound (CRLB), while for LR it is asymptotically lower-bounded by \cite{Cerutti2020}: 

\begin{equation}\label{eq:MSEbound_no_whitening}
\begin{split}
\mathrm{MSE_b} & = \mathrm{tr}\left(\boldsymbol{\Pi}\{\widehat{r}\} \,\mathbf{C}\, \boldsymbol{\Pi}^{\mathrm{H}}\{\widehat{r}\}\right) + \\ & + \mathrm{tr}\left( \Delta \boldsymbol{\Pi}\{\widehat{r}\} \, \widetilde{\mathbf{R}} \, \Delta\boldsymbol{\Pi}^{\mathrm{H}}\{\widehat{r}\}\right)
\end{split}
\end{equation}

where: \textit{(i)} $\boldsymbol{\Pi}\{\widehat{r}\}$ is the asymptotic ($L \rightarrow \infty$) (true) position-dependent matrix $\boldsymbol{\Pi}(\mathbf{\bar{p}},\bar{\theta})$ evaluated for the estimated rank, either $\widehat{r}$ (for JS-LR) or $\widehat{r}_T$ and $\widehat{r}_R$ (for DS-LR); \textit{(ii)} $\Delta \boldsymbol{\Pi}\{\widehat{r}\} = \boldsymbol{\Pi}\{\widetilde{r}\} - \boldsymbol{\Pi}\{\widehat{r}\}$ is the difference between the asymptotic position-dependent matrix computed for the true channel rank ($\widetilde{r}$ for JS-LR or $\widetilde{r}_T$ and $\widetilde{r}_R$ for DS-LR) and for the estimated one. Therefore, the first term accounts for the residual noise contribution, while the second for misparameterization (errors in the estimated diversity orders of the channel). Here, we estimate the channel rank from the correlations' eigenvalues, in descending order, according to the $99.9$ percentile of their cumulative sum.

\subsection{S1 (multipath scenario far from the BS)}\label{subsect:S1}

\begin{figure}[b!]
    \centering
    \subfloat[]{\includegraphics[width=0.48\columnwidth]{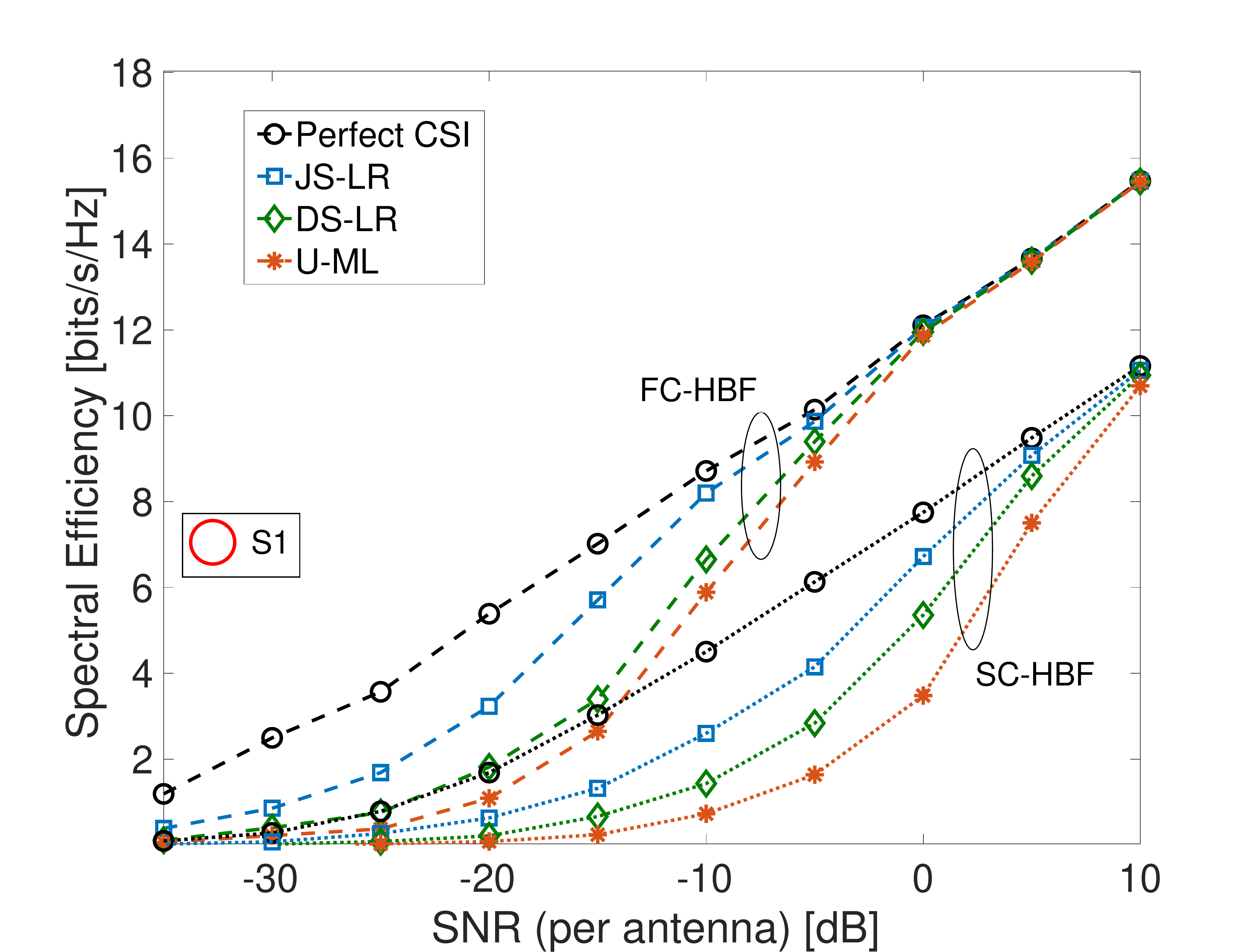}\label{subfig:SEvsSNR_FC_SC}}\hspace{0.2mm}
    \subfloat[]{\includegraphics[width=0.48\columnwidth]{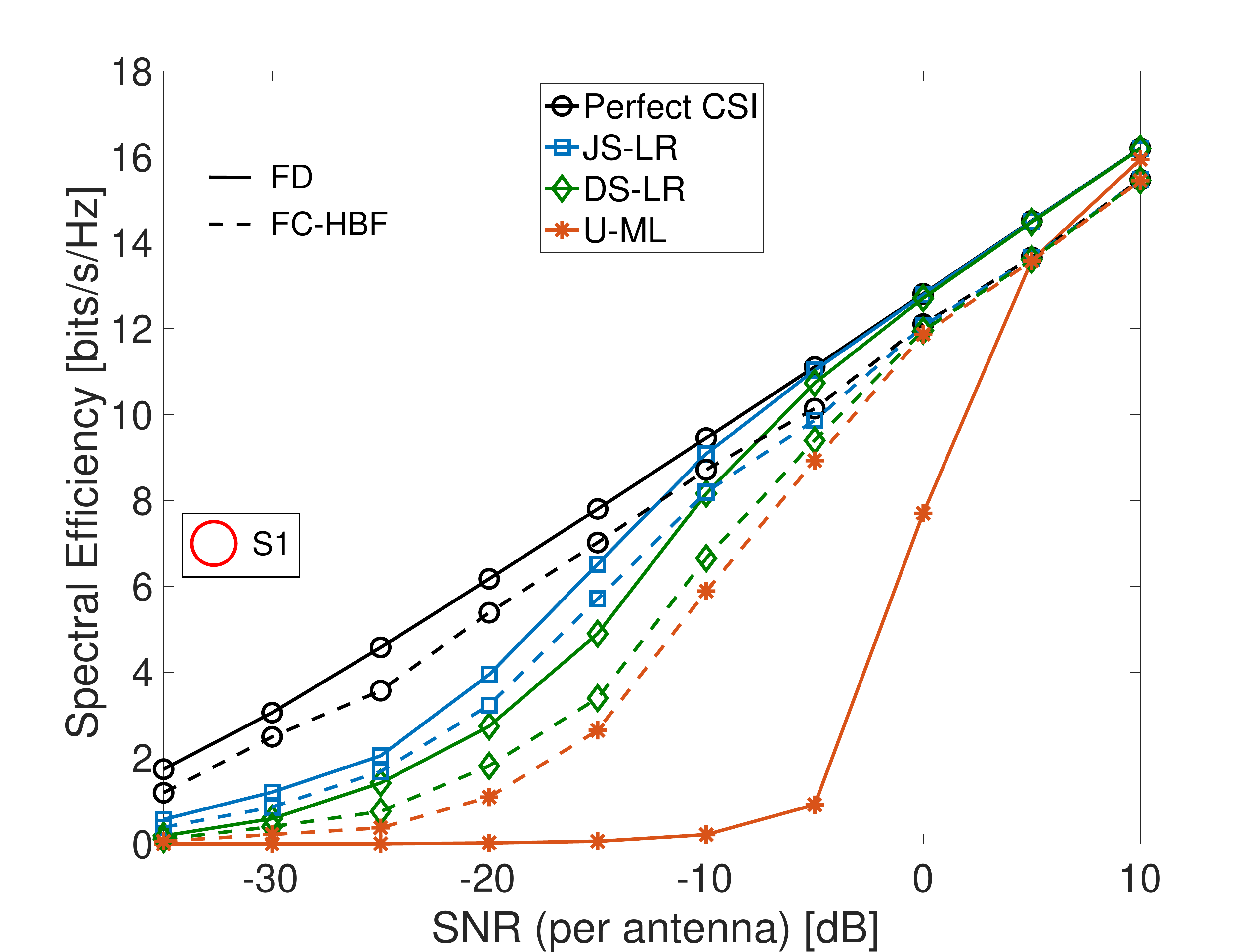}\label{subfig:SEvsSNR_FD_FC}}\\
    \caption{Spectral efficiency of FC-HBF vs. SC-HBF (\ref{subfig:SEvsSNR_FC_SC}) and FC-HBF vs. FD (\ref{subfig:SEvsSNR_FD_FC}), with U-ML, JS-LR, and DS-LR channel estimation methods and perfect CSI; $L = 1000$ training blocks (vehicle passages)}
    \label{fig:SEvsSNR}
\end{figure}

\begin{figure}[!t]
    \centering
    \includegraphics[width=0.48\columnwidth]{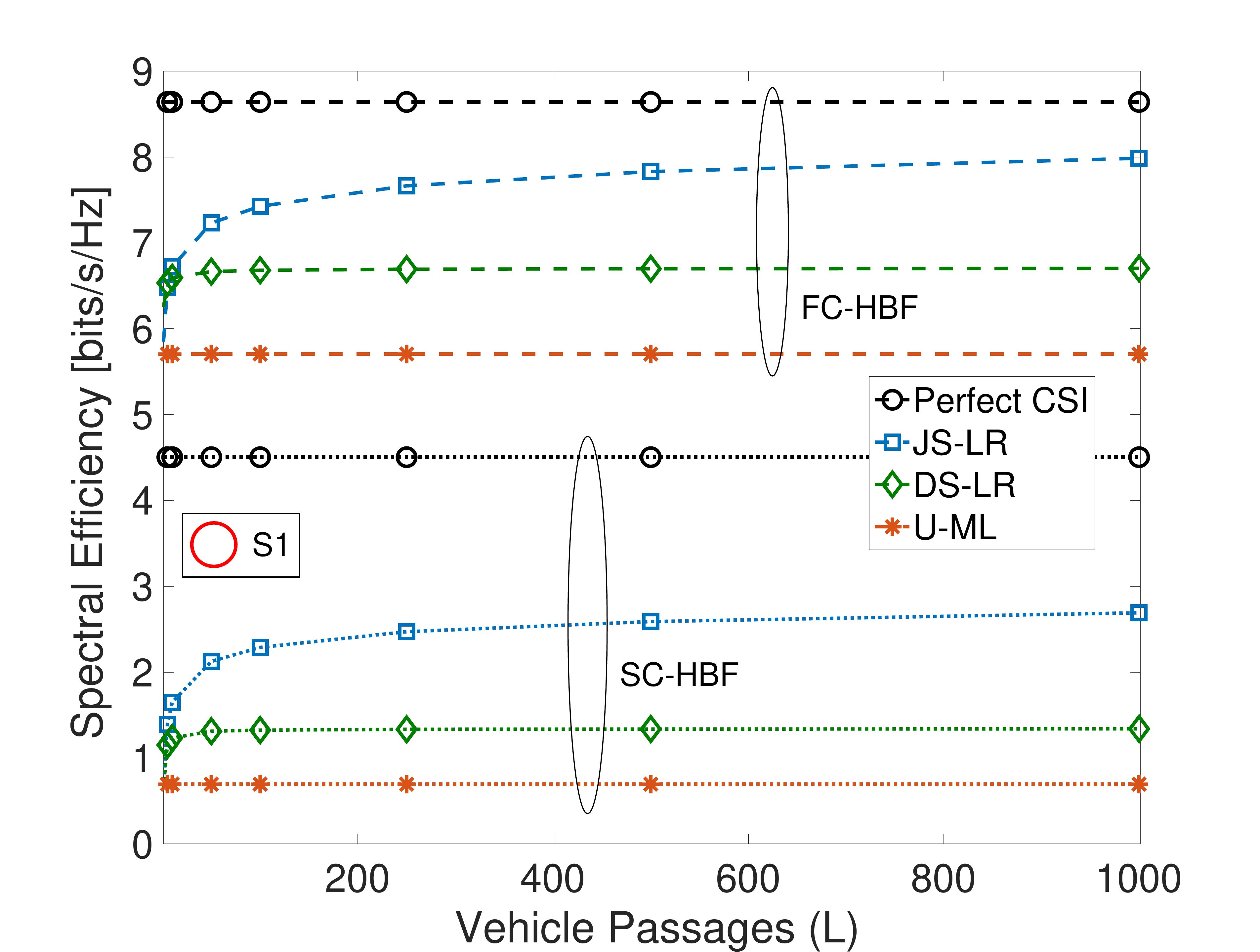}
    \caption{Spectral efficiency of FC-HBF and SC-HBF systems varying the number of vehicle passages $L$}
    \label{fig:SEvsL}
\end{figure}

\begin{figure}[t!]
    \centering
    \subfloat[FC-HBF]{\includegraphics[width=0.48\columnwidth]{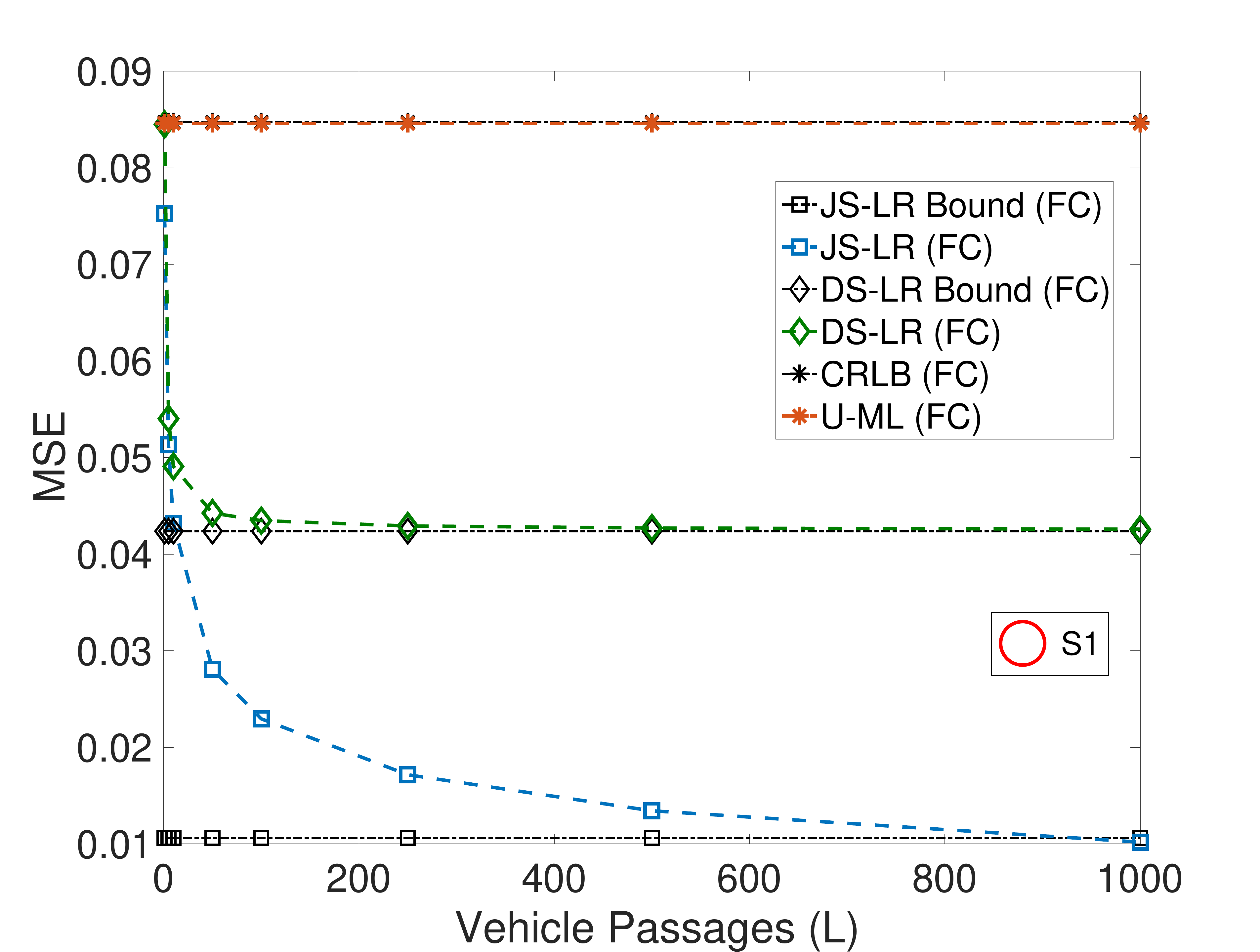}\label{subfig:MSEvsL_FC}}\hspace{0.2mm}
    \subfloat[SC-HBF]{\includegraphics[width=0.48\columnwidth]{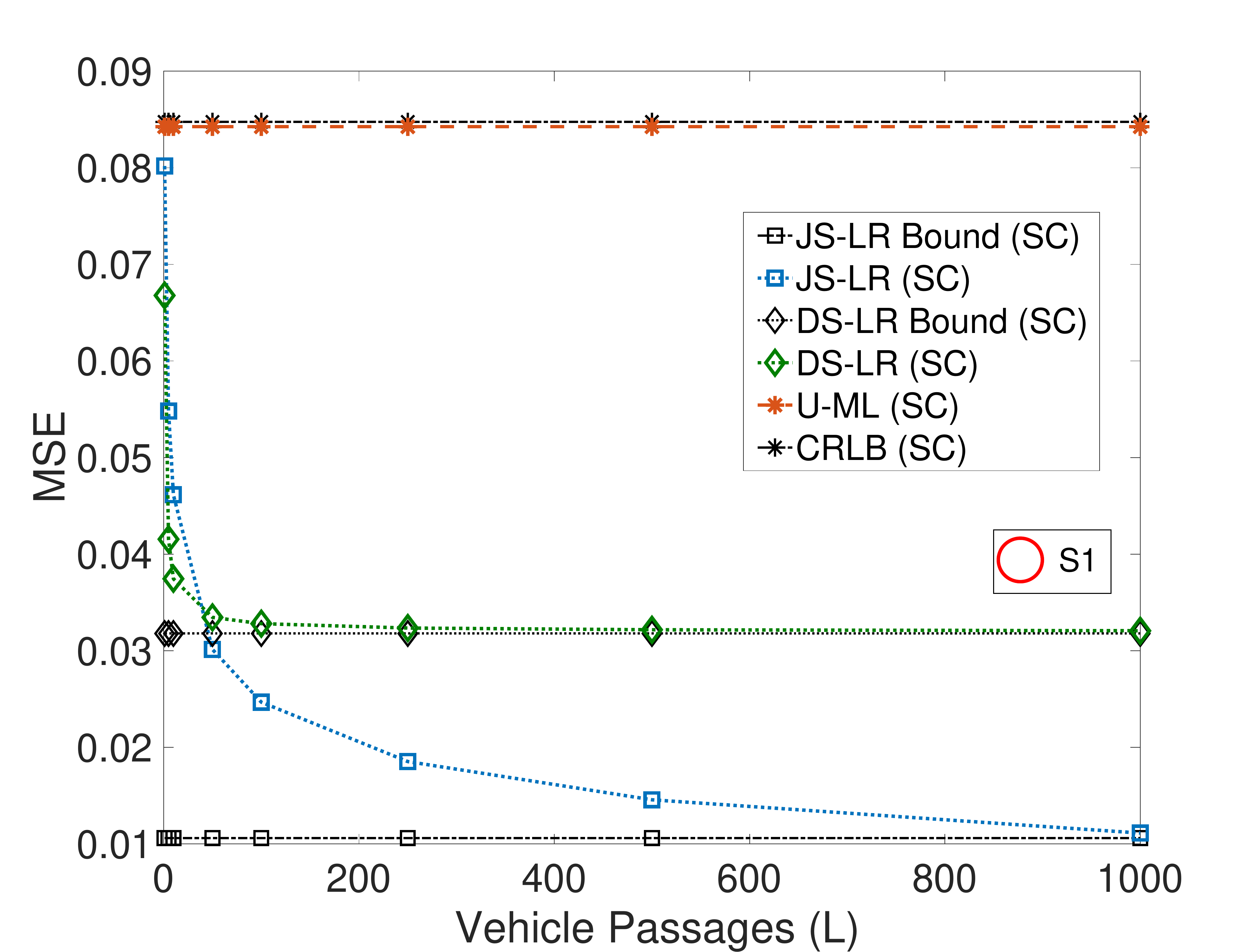}\label{subfig:MSEvsL_SC}}\\
    \caption{MSE on compressed channel estimation varying the number of vehicle passages $L$ for FC-HBF (\ref{subfig:MSEvsL_FC}) and SC-HBF (\ref{subfig:MSEvsL_SC}) systems, with U-ML, JS-LR, and DS-LR methods and corresponding theoretical bounds; SNR (per antenna) = $-10$ dB}
    \label{fig:MSEvsL}
\end{figure}

\begin{figure}[t!]
    \centering
    \includegraphics[width=0.48\columnwidth]{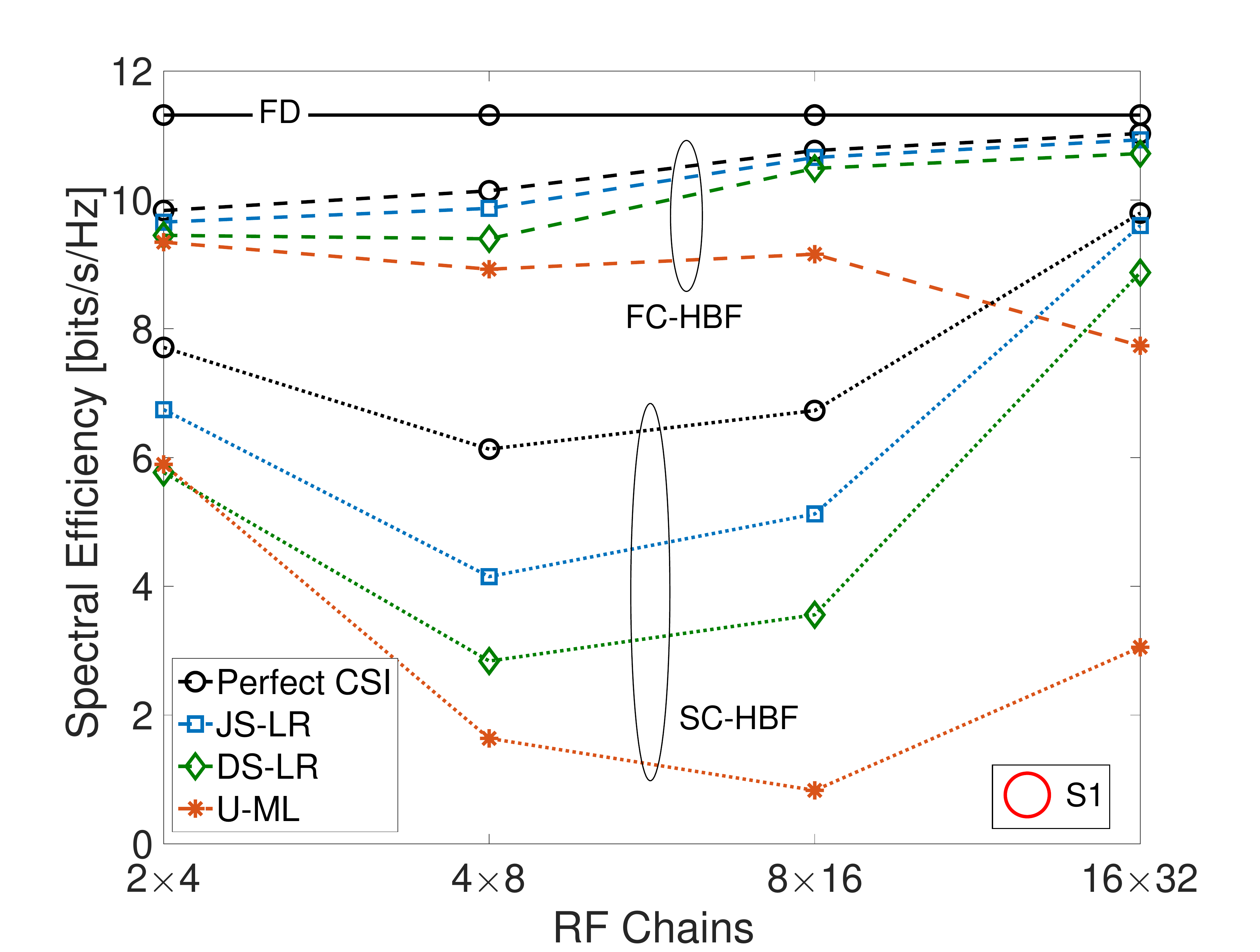}
    \caption{Spectral efficiency of FC-HBF and SC-HBF systems varying the number of RF chains ($N_T^{RF} \times N_R^{RF}$), with U-ML, JS-LR, and DS-LR channel estimation methods and perfect CSI; $L = 1000$ vehicle passages, SNR = $-10$ dB (FD as upper bound) }
    \label{fig:SEvsNRF}
\end{figure}
\begin{figure}[t!]
    \centering
    \subfloat[FC-HBF]{\includegraphics[width=0.48\columnwidth]{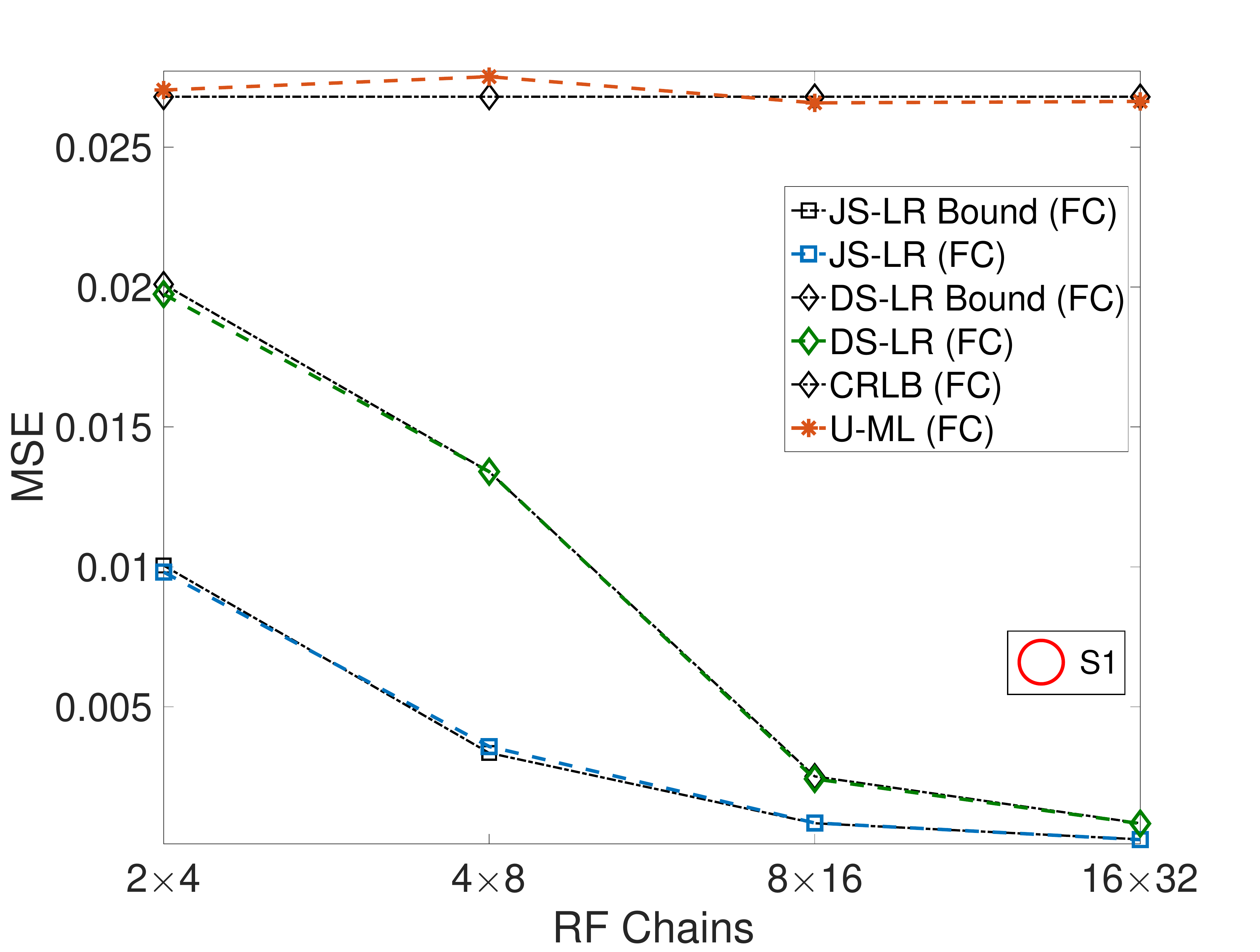}\label{subfig:MSEvsNRF_FC}} \hspace{0.2mm}
    \subfloat[SC-HBF]{\includegraphics[width=0.48\columnwidth]{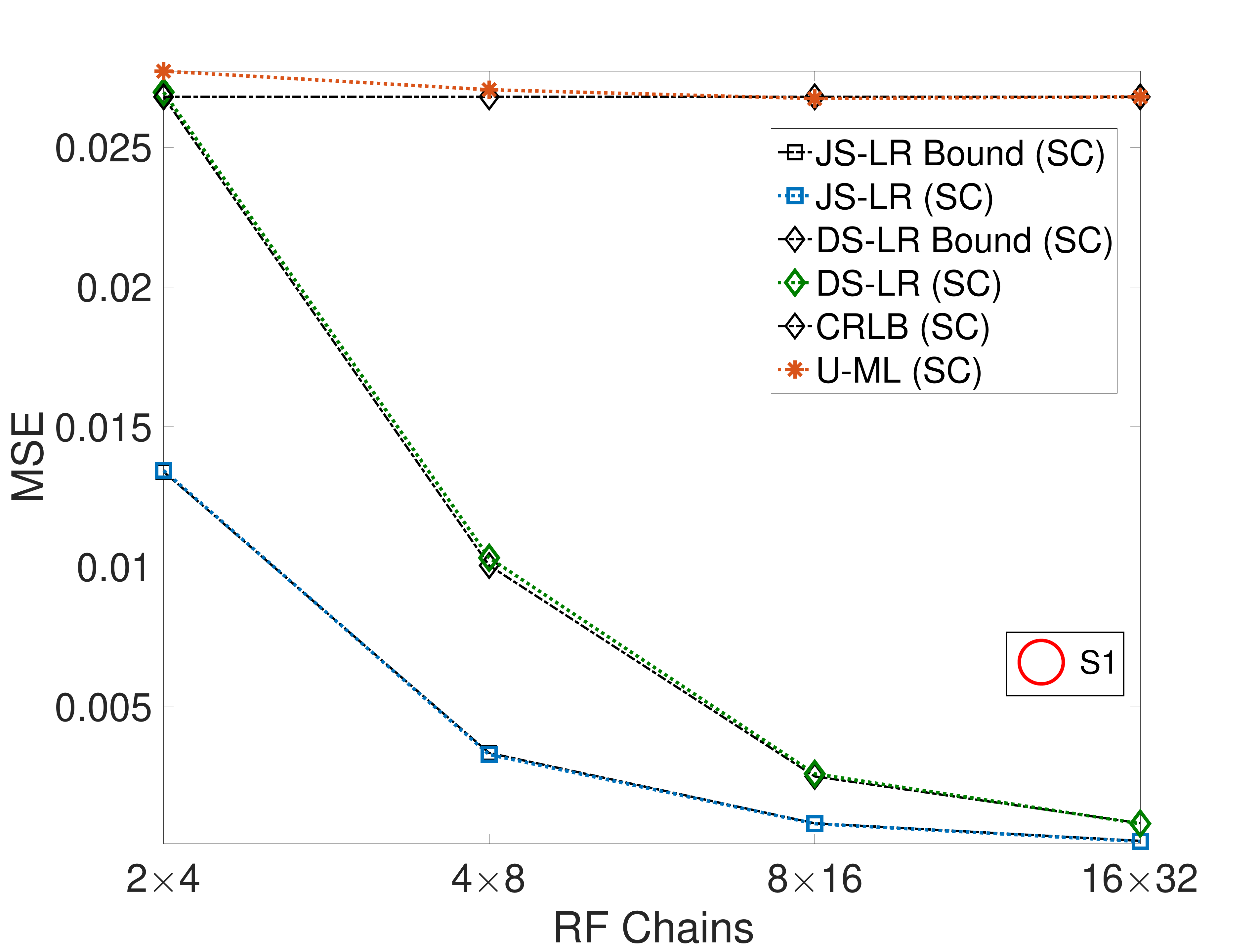}\label{subfig:MSEvsNRF_SC}}\\ 
    \caption{MSE on compressed channel estimation varying the number of RF chains ($N_T^{RF} \times N_R^{RF}$) for FC-HBF (\ref{subfig:MSEvsNRF_FC}) and SC-HBF (\ref{subfig:MSEvsNRF_SC}) systems, with U-ML, JS-LR, and DS-LR channel estimation methods and corresponding theoretical bounds; $L = 1000$ vehicle passages, SNR = $-10$ dB }
    \label{fig:MSEvsNRF}
\end{figure}

Fig. \ref{fig:SEvsSNR} shows the achievable SE in \eqref{eq:SE} varying the SNR per antenna, i.e., before analog beamforming, for FC-HBF vs. SC-HBF  architectures (Fig. \ref{subfig:SEvsSNR_FC_SC}) and FC-HBF vs. FD (Fig. \ref{subfig:SEvsSNR_FD_FC}). The SE is evaluated with four different degrees of channel knowledge: \textit{(i)} perfect Channel State Information (CSI) (black lines); \textit{(ii)} optimal JS-LR channel estimation (blue lines); \textit{(iii)} sub-optimal DS-LR channel estimation (green lines); \textit{(iv)} U-ML channel estimation (red lines). The FD performance are computed as benchmark, with precoders and combiners obtained with \eqref{eq:DigitalPrecoderDesign} and \eqref{eq:DigitalCombinerDesign}, respectively, by using the U-ML-, JS-LR- and DS-LR-estimated full channel $\widehat{\mathbf{H}}$. In all the LR implementations, the number of training vehicle passages is $L=1000$. 
As expected, the SC-HBF architectures provides worst performance compared to FC-HBF, as a consequence of the reduced analog beamforming gain. In both FC-HBF and SC-HBF configurations, however, we notice the remarkable performance gain compared to U-ML provided by DS-LR and especially JS-LR. For FC-HBF, at a reference SNR of $-10$ dB, the SE gap amounts to 0.8 bits/s/Hz for DS-LR and to 2.6 bits/s/Hz for JS-LR (Fig. \ref{subfig:SEvsSNR_FC_SC}). It can be appreciated that, for FC-HBF system, the use of JS-LR channel estimation method allows to practically approach the perfect CSI case ($R\approx8$ bits/s/Hz) with 5 dB less of SNR compared to U-ML. For SC-HBF, instead, the DS-LR provides a SE gain of 0.7 bits/s/Hz and JS-LR 1.9 bits/s/Hz, while the SNR gain is even higher, up to $\approx 10$ dB for $R\approx3$ bits/s/Hz, while the perfect CSI case is attained for $SNR \geq 10$ dB. As can be observed from Fig. \ref{subfig:SEvsSNR_FD_FC}, the FC-HBF (dashed lines) performance practically matches the FD one (solid lines), apart from a negligible SE penalty due to the fixed spatial sampling provided by the use of analog codebooks. According to Subsection \ref{subsect:lossylossless}, this result is expected, as $N_T^{RF}=4>r_T=3$ and $N_R^{RF}=8>r_R=3$. 

Figs. \ref{fig:SEvsL} and \ref{fig:MSEvsL} depict, respectively, the behavior of the SE and of the MSE of both FC-HBF and SC-HBF systems with respect to the number of vehicle passages $L$, for SNR = $-10$ dB. The DS-LR method requires a lower number of passages, approximately $L=50$ for HBF, to converge to its asymptotic MSE bound, whereas for JS-LR method the convergence is guaranteed for $L=500$ blocks (HBF). It is important to emphasize that, at the cost of approximately 1.5 bits/s/Hz in SE (Fig. \ref{fig:SEvsL}), we have a remarkable gain in complexity, which is approximately ruled by the computation of the eigenvectors of correlation matrices in \eqref{eq:STchannel_vector_correlation_sample} and \eqref{eq:SST_compressed_channel_correlations_sample_1}-\eqref{eq:SST_compressed_channel_correlations_sample_2}, since the computational cost of $\mathrm{eig} (\hat{\mathbf{R}})$ is $\mathcal{O}((N^{RF}_T N^{RF}_R)^3) \geq \mathcal{O}((N^{RF}_T)^3) + \mathcal{O}((N^{RF}_R)^3)$, as required for $\mathrm{eig} (\hat{\mathbf{R}}_{T})$ and $\mathrm{eig} (\hat{\mathbf{R}}_{R})$. In general, the results show a significant performance gain with LR compared to U-ML on whole SNR range.

The last results on S1 are related to the SE and MSE performance of HBF varying the number of RF chains $N_R^{RF}\times N_T^{RF}$, summarized in Figs. \ref{fig:SEvsNRF} and \ref{fig:MSEvsNRF}, fixing $L = 1000$ vehicle passages and SNR per antenna of $-5$ dB. 
For FC-HBF systems, the SE gap between LR and U-ML goes proportionally to the number of RF chains. The MSE in Fig. \ref{subfig:MSEvsNRF_FC} explains the SE performance: by increasing $N_R^{RF}$ and $N_T^{RF}$, the MSE of LR decreases with the increasing sparsity of the compressed channel; conversely, the MSE of U-ML does not change. For SC-HBF architectures, instead, Fig. \ref{fig:SEvsNRF} shows an interesting trade-off between having a high analog resolution (few RF chains) or having a high digital resolution (i.e., approach the FD system, for $16 \times 32$ RF chains). A high analog resolution implies a comparably low LR gain with respect to U-ML, as the compressed channel sparsity decreases; a high digital resolution leads to a significant sparsity of $\widetilde{\mathbf{H}}$ and thus to a huge LR gain. For any HBF configuration in between, the performance decreases. The MSE in \ref{subfig:MSEvsNRF_SC} exhibits a similar trend with respect to Fig. \ref{subfig:MSEvsNRF_FC}, again explaining the SE gain of LR compared to U-ML.

\subsection{S2 (single-path scenario close to the BS)}\label{subsect:S2}
\begin{figure}[!tb]
    \centering
    \subfloat[]{\includegraphics[width=0.48\columnwidth]{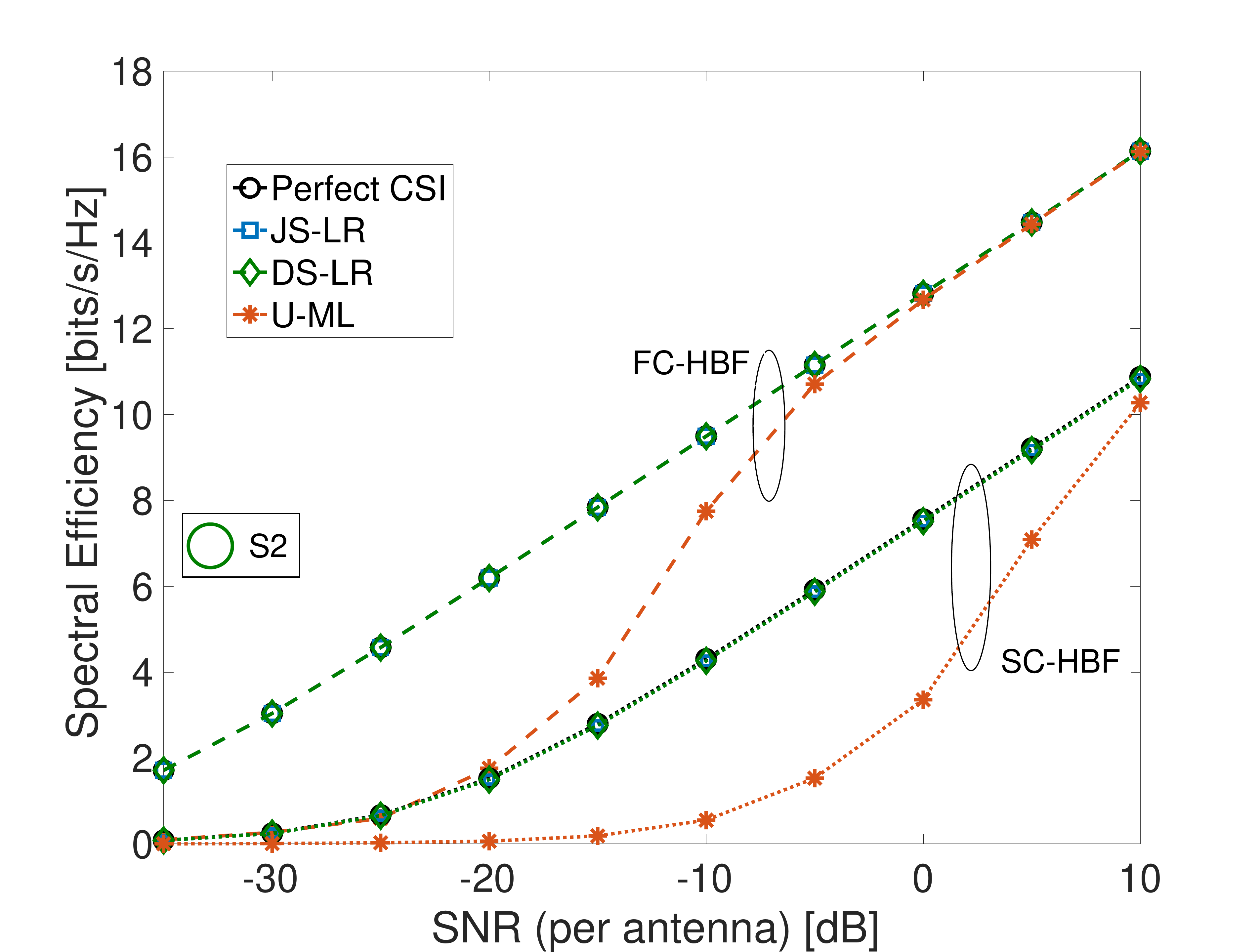}\label{subfig:SEvsSNR_FC_SC_vicino}} \hspace{0.2mm}
    \subfloat[]{\includegraphics[width=0.48\columnwidth]{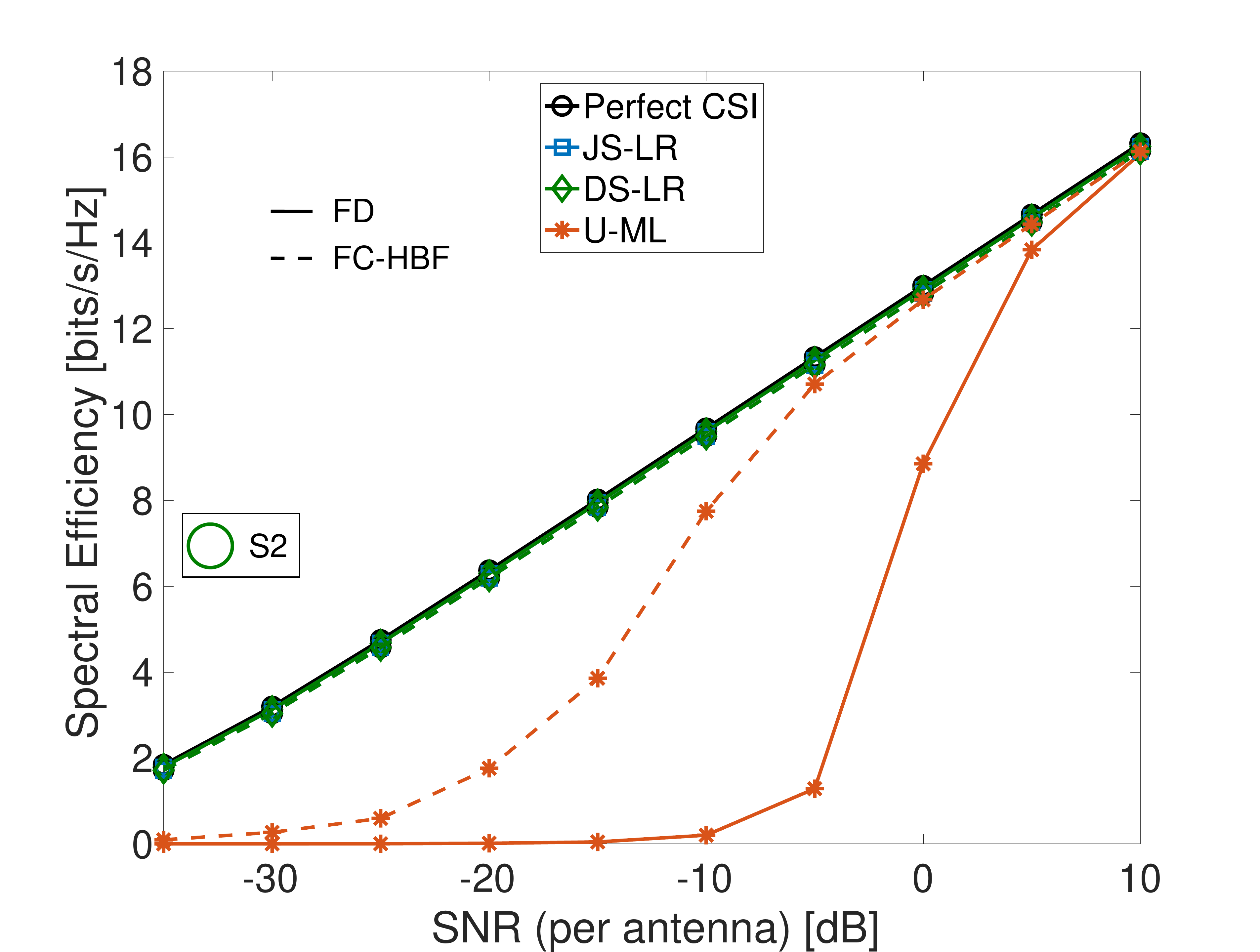}\label{subfig:SEvsSNR_FD_FC_vicino}}
    \caption{Spectral efficiency of FC-HBF vs. SC-HBF (\ref{subfig:SEvsSNR_FC_SC_vicino}) and FC-HBF vs. FD (\ref{subfig:SEvsSNR_FD_FC_vicino}), with U-ML, JS-LR, and DS-LR channel estimation methods and perfect CSI; $L = 1000$ training blocks (vehicle passages)}
    \label{fig:SEvsSNR_vicino}
\end{figure}

\begin{figure}[!tb]
    \centering
    \includegraphics[width=0.48\columnwidth]{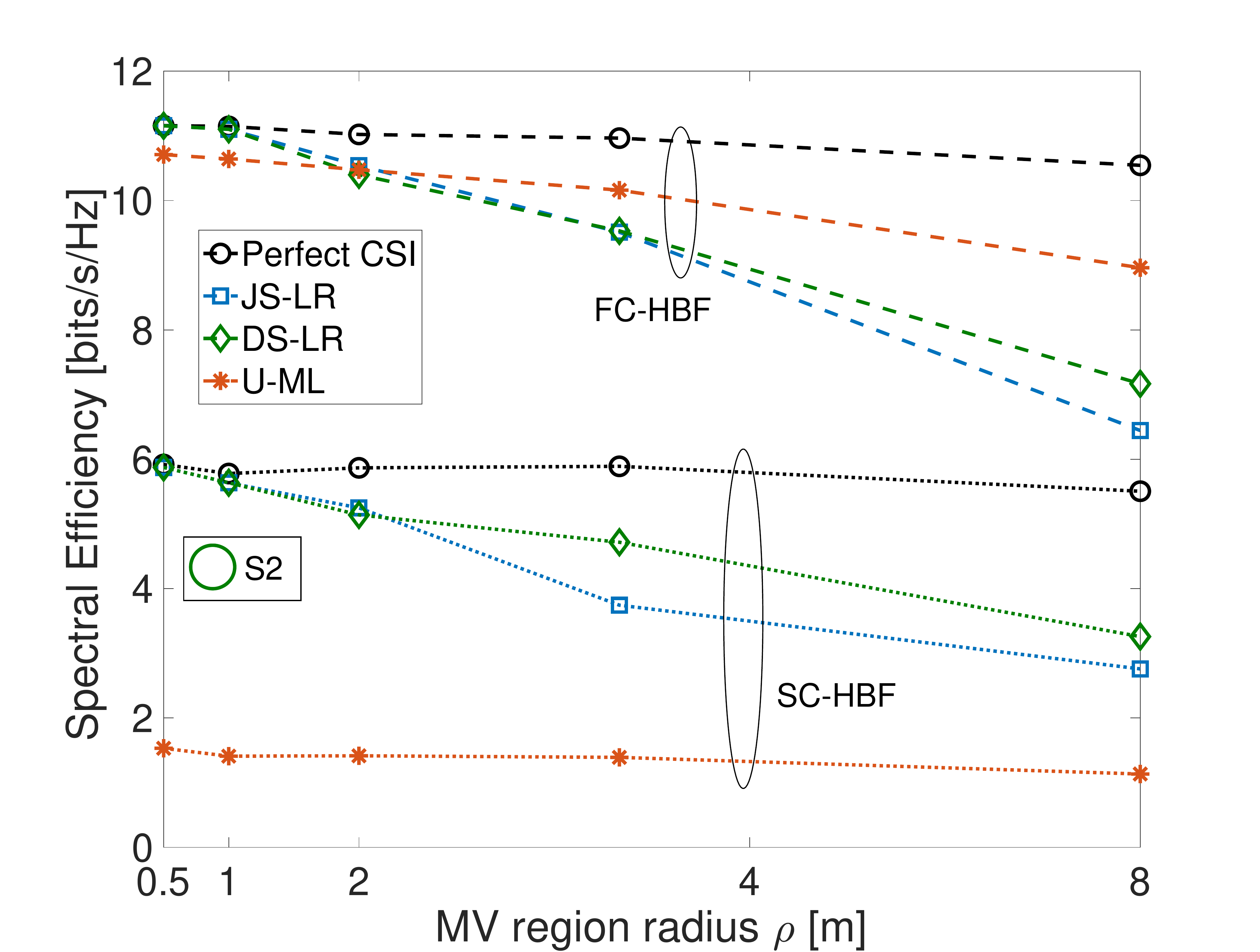}
    \caption{Spectral efficiency of FC-HBF vs. SC-HBF varying the MV region radius $\rho$, with U-ML, JS-LR, and DS-LR channel estimation methods and perfect CSI; $L = 1000$ training blocks (vehicle passages), SNR = $-5$ dB}
    \label{fig:SEvsROI_FC_SC_vicino}
\end{figure}

\begin{figure}[!tb]
    \centering
    \includegraphics[width=0.48\columnwidth]{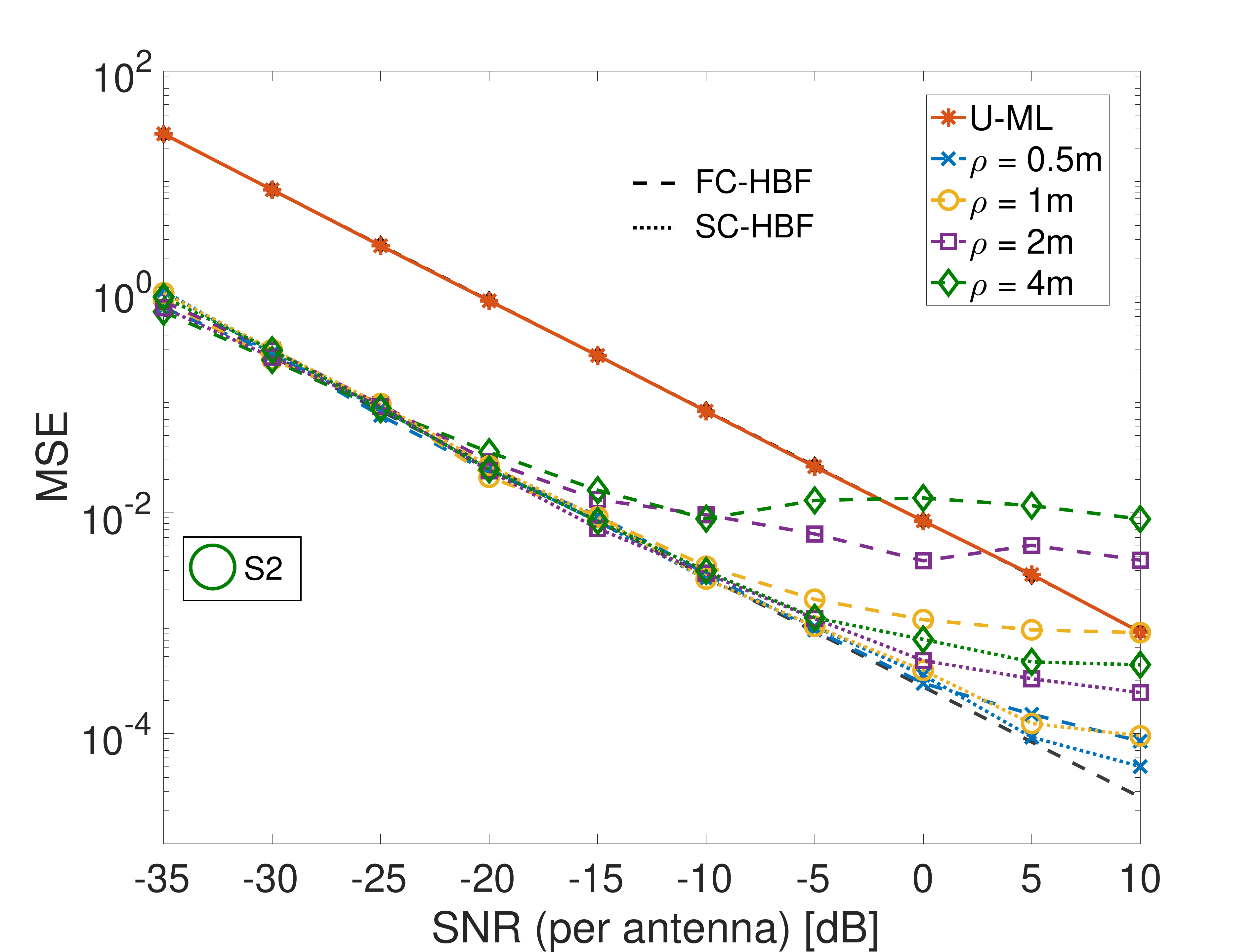}
    \caption{MSE on JS-LR channel estimation of FC-HBF vs. SC-HBF, varying the SNR, function of the MV region radius $\rho$; $L = 1000$ training blocks (vehicle passages), SNR = $-5$ dB }
    \label{fig:MSEJSvsROI_FC_SC_vicino}
\end{figure}

The last set of results are related to the single-path scenario S2 (Fig. \ref{fig:street}). Similarly to S1, we report in Fig. \ref{fig:SEvsSNR_vicino} the SE varying the SNR before beamforming for FC-HBF vs. SC-HBF  architectures (Fig. \ref{subfig:SEvsSNR_FC_SC_vicino}) and FC-HBF vs. FD (Fig. \ref{subfig:SEvsSNR_FD_FC_vicino}). Again, we consider U-ML, JS-LR and DS-LR channel estimation methods, and the perfect CSI case as upper bound, with $L=1000$ vehicle passages. Compared to the multipath scenario S1, JS-LR and DS-LR channel estimation methods approach the perfect CSI case, for both FC-HBF and SC-HBF. This can be explained by considering that for a single-path channel, the sparsity degrees in \eqref{eq:JS_sparsity} and \eqref{eq:DS_sparsity_1}-\eqref{eq:DS_sparsity_2} are maximum, and the residual error on the LR-estimated channel does not remarkably impact on the SE. 

In S2, the proposed system performance is more sensitive to the MV region size compared to S1. Fig. \ref{fig:SEvsROI_FC_SC_vicino} shows the SE of FC-HBF and SC-HBF for all the channel estimation methods varying $\rho$ (MV region radius) from 0.5 m to 4 m (the latter basically considering the whole area of the crossing in Fig. \ref{fig:street}), for a fixed SNR = $-5$ dB. We notice that the JS-LR and DS-LR performance drastically decrease with $\rho$, especially for FC-HBF, and can be even worse than the U-ML one. This is a direct consequence of the AoDs/AoAs variation within the selected MV region, which exceeds the system resolution (spatial selectivity of BS array) and leads to subspace decorrelation. In other words, the ensemble of received sequences $\{\overline{\mathbf{y}}\}_{\ell=1}^L$ (Subsection \ref{subsect:LR}) do not have the same propagation subspace. As the spatial resolution of FC-HBF systems is higher of SC-HBF one, the effect for the former is stronger. This is further confirmed by the MSE of JS-LR varying the SNR before beamforming (Fig. \ref{fig:MSEJSvsROI_FC_SC_vicino}), where we notice a progressive deviation from the asymptotic MSE bound (black, dashed line), proportional to $\rho$ and to the SNR. For low SNRs, the imperfect modal filtering provided by an excessive cluster size $\rho$ is negligible for low SNRs, where the noise is dominant, while is relevant for higher SNR values. This last result shows a tradeoff between the performance of the HBF system and the available resolution in MS position accuracy, which allows to set up the MV regions. FC-HBF systems provide superior performance, attaining FD one, but require a very accurate positioning, while SC-HBF allows to relax this constraint.

\section{Conclusion}\label{sect:conclusion}

In this paper, we propose a training-based multi-stage channel estimation method for hybrid mmWave/sub-THz MIMO systems, based on terminals in mobility (e.g., in V2I or V2N scenarios), in the 5G and future 6G context. The first training stage relies on a novel multi-vehicular codebook-based beam alignment procedure to obtain the optimal analog precoder and combiner. In the second training stage, we adapt the Low-Rank (LR) channel estimation to hybrid systems, and we propose two LR methods, namely Joint-Space Low-Rank (JS-LR) and Disjoint-Space Low-Rank (DS-LR), for deriving the hybrid channel eigenmodes. Finally, in the last stage, i.e., communications phase, we derive the digital precoders/combiners based on both the optimal analog precoder/combiner pair from the first stage and the hybrid channel eigenmodes from the second stage.

The proposed LR methods are analyzed numerically, but realistically, considering a V2I/V2N urban scenario based on OpenStreetMap for roads/buildings topology and SUMO for the vehicular mobility. The channel is generated by ray-tracing and the performances are compared in terms of Spectral Efficiency (SE) and Mean Squared Error (MSE) on channel estimation. The metrics on Full Digital (FD) system are reported as benchmark, as well as the performance of the Unconstrained Maximum Likelihood (U-ML). The two proposed solutions, i.e., optimal (JS-LR) and sub-optimal (DS-LR), are examined for both Fully Connected (FC-HBF) and Sub-Connected (SC-HBF) architectures varying SNR, training vehicles' number, RF chains configuration, and channel configurations, i.e., multipath (S1) and single-path (S2).

The achieved results proved the great advantage of our solution. In particular, we observed that in the single-path scenario (S2), both JS-LR and DS-LR solutions attain the SE of the perfect CSI results. Moreover, the FC-HBF architecture exhibits similar performance to the benchmark (FD). In the multipath scenario (S1), both solutions show better performance compared to the U-ML estimator and attain the perfect CSI for an SNR $> -5$ dB for FC-HBF and SNR $> 10$ dB for SC-HBF. In general, we can conclude that, under the same conditions, the FC-HBF architectures perform better than SC-HBF in terms of SE and present the same MSE. However, the SC-HBF architectures are less sensitive to positioning errors, which impacts on the size of the multi-vehicular region used for training. 

Another aspect of interest is that, as the number of RF chains increases, the performance gap (LR-U-ML) of the FC-HBF architectures increases, while for SC-HBF architecture, we observe that pursuing a trade-off between digital and analog resolution is detrimental, and it is more appropriate to consider a system with high digital resolution (higher number of RF chains), or high analog resolution (low number of RF chains), with the former being preferable. Concerning the comparison between JS-LR and DS-LR, we found that, under the same conditions, the former shows a better SE and MSE. Moreover, the SE gap is greater especially in the multipath and/or low SNR scenario, and it reduces in the single-path and/or high SNR scenario. Consequently, in these cases, the DS-LR method is recommended as it is significantly less complex, requiring less training vehicles for convergence to the theoretical bound. In real cases, the presence of neighbouring vehicles (even parked) is expected to make the estimates to be time-varying, but nevertheless the BS can always command the MSs to repeat some MV learning steps for refinement of the position-based estimate.

\section*{Acknowledgment} \label{app:acknowledgment}

The research has been carried out in the framework of the Huawei-Politecnico di Milano Joint Research Lab. The Authors want to acknowledge Huawei Milan Research Center.

\bibliographystyle{IEEEtran}
\bibliography{myBib}

\end{document}